\documentclass[%
 twocolumn,
 amsmath,amssymb,
 aps,
prb,
]{revtex4-2}

\usepackage{graphicx}
\graphicspath{{figures/}}
\usepackage{dcolumn}
\usepackage{bm}
\usepackage{hyperref}
\usepackage{xspace}
\newcolumntype{C}{>{$}c<{$}}

\newcommand{\qr}{Q_{\textrm{R}}\xspace}
\newcommand{\qs}{Q_{\textrm{S}}\xspace}
\newcommand{\qir}{Q_{\textrm{IR}}\xspace}
\newcommand{\qlz}{Q_{\textrm{LZ}}\xspace}
\newcommand{\qhx}{Q_{\textrm{HX}}\xspace}
\newcommand{\qlx}{Q_{\textrm{LX}}\xspace}
\newcommand{\olz}{\Omega_{\textrm{LZ}}\xspace}
\newcommand{\olx}{\Omega_{\textrm{LX}}\xspace}
\newcommand{\ohx}{\Omega_{\textrm{HX}}\xspace}

\usepackage{xpunctuate}
\DeclareRobustCommand\etal{\xperiodafter{\emph{et al}}}

\begin{document}

\title{Light-induced translation symmetry breaking via nonlinear phononics}

\author{Adri\'an G\'omez Pueyo}
\email{adrian.gomez@polytechnique.edu}
\affiliation{
  CPHT, CNRS, Ecole Polytechnique, IP Paris, F-91128 Palaiseau, France
}%

\author{Alaska Subedi}
\email{alaska.subedi@polytechnique.edu}
\affiliation{
  CPHT, CNRS, Ecole Polytechnique, IP Paris, F-91128 Palaiseau, France
}%

\date{\today}

\begin{abstract}
  Light has a wavelength that is usually longer than the size of the unit cell
  of crystals. Hence, even intense light pulses are not expected to break the
  translation   symmetry of materials.  However, certain materials, including
  KTaO$_3$, exhibit   peaks in their Raman spectra corresponding to their Brillouin
  zone boundary phonons due to second-order Raman processes, which provide a
  mechanism to drive these phonons using intense midinfrared lasers.  We
  investigated the possibility of breaking the translation symmetry of KTaO$_3$ by
  driving its highest-frequency transverse optic mode $Q_{\textrm{HX}}$ at the $X$
  $(0,\frac{1}{2},0)$ point.  Our first principles calculations show that the 
  energy curve of the transverse acoustic mode $Q_{\textrm{LZ}}$ at $X$ softens 
  and develops a double-well   shape as the value of the $Q_{\textrm{HX}}$
  coordinate is increased, while that of the other transverse acoustic component
  $Q_{\textrm{LX}}$ hardens when the value of the $Q_{\textrm{HX}}$ coordinate is
  similarly varied. We performed similar total energy calculations as a function 
  of the $Q_{\textrm{HX}}$ coordinate and electric field to extract the nonlinear 
  coupling between them. These were then used to construct the coupled equations 
  of motion for the three phonon coordinates in the presence of an external pump
  term on the $Q_{\textrm{HX}}$ mode, which we numerically solved for a range of
  pump frequencies and amplitudes.  We find that 465 MV/cm is the smallest pump 
  amplitude that leads to an oscillation of the $Q_{\textrm{LZ}}$ mode at a
  displaced position, hence, breaking the translation symmetry of the material. 
  Such highly intense light pulses cannot be generate by currently available 
  laser sources, and they have the possibility to damage the material.
  Nevertheless, our work shows that light can in principle be used to break the
  translation symmetry of a material via nonlinear phononics.
\end{abstract}

\maketitle


\section{Introduction}
\label{sec:introduction}

Ultrafast structural control of materials by coherently exciting their
phonons using intense laser pulses is an active area of research
\cite{Mankowsky2016, Salen2019, Subedi2021}.  This field of nonlinear phononics 
started when F\"orst \etal realized that a $\qs\qir^2$ nonlinear
coupling between fully-symmetric Raman $\qs$ and infrared
$\qir$ phonon modes can cause a displacement of the lattice along the $\qs$ 
coordinate when the $\qir$
mode is externally pumped~\cite{Forst2011}. A limitation of this type 
of coupling is the inability to break any crystal symmetry of a material. 
Nevertheless, an investigation of this nonlinearity in perovskite 
ferroelectrics using first principles calculations has found that this 
mechanism can be used to switch their electrical 
polarization~\cite{Subedi2015},  and this theoretical prediction has 
been partially confirmed in subsequent
experiments~\cite{Mankowsky2017, Henstridge2022}.

Historically, only cubic nonlinearities between Raman and
infrared phonons were investigated in the context of ionic
Raman scattering~\cite{Wallis1971, Martin1974}. However, first 
principles calculations in
Ref.~\cite{Subedi2014} showed that a symmetry-breaking Raman
phonon mode $\qr$ can have substantial quartic-order $\qr^2\qir^2$ coupling
with an infrared phonon mode.  Such a large quartic-order coupling
between two infrared modes has also been calculated in oxide
paraelectrics, which has been used to predict light-induced
ferroelectricity~\cite{Subedi2017}. Radaelli has shown that 
driving degenerate infrared modes along orthogonal directions 
can cause displacement of the lattice along a symmetry-breaking Raman 
mode due to a cubic-order nonlinearity~\cite{Radaelli2018}, while a separate 
study has shown that the symmetry-breaking Raman mode oscillates about 
the equilibrium position with the difference frequency when nondegenerate 
infrared phonons are driven along orthogonal 
directions~\cite{Juraschek2017}. Additional	theoretical and experimental 
studies have demonstrated that nonlinear phononics 
is a useful technique to control the crystal structure and,	hence, the 
physical properties of materials~\cite{Fechner2016, Gu2016, Gu2017,
Juraschek2017b, Fechner2018, Gu2018, Khalsa2018, Park2019, Juraschek2021,
Kaaret2021, Feng2022,
Nova2017, Hortensius2020,Neugebauer2021, Afanasiev2021, Disa2020,
Melnikov2020, Stupakiewicz2021, Disa2021}.  
However, these studies have only focused on light-induced structural
modifications that do not change the size of the unit cell thus far.

In this paper, we investigate the possibility of breaking the
translation symmetry of KTaO$_3$ using light by driving its Brillouin 
zone boundary phonon modes.  This was motivated by the observation of
large two-phonon peaks due to zone boundary modes in the Raman
spectrum of this material \cite{Nilsen1967}, indicating that these 
modes couple significantly to light.  We obtained the nonlinear 
couplings between the highest-frequency transverse optic (TO) mode 
$\qhx$ and doubly-degenerate components of the
transverse acoustic (TA) mode $\qlz$ and $\qlx$ at the 
$X$ $(0,\frac{1}{2},0)$ point using first principles total-energy 
calculations, which show that the TA $\qlz$ mode softens when
the orthogonal TO coordinate $\qhx$ has a finite value.  The coupling 
between the $\qhx$ mode and electric field was obtained from similar
total energy calculations.  These were then used to construct coupled
equations of motion for the phonon coordinates.  Their numerical
solutions showed that the TA $\qlz$ mode can rectify and break the
translation symmetry of the lattice when the $\qhx$ mode is pumped.
However, 465 MV/cm is the lowest pump amplitude that causes the
rectification.  Beyond the possibility of sample damage by such an
intense pulse, the required intensity is also at least an order of
magnitude larger than that can be produced by currently available
midinfrared laser sources.  Nevertheless, our study demonstrates that
light can in principle be used to break the translation symmetry of crystals
through nonlinear phononics and motivates search for materials that
exhibit large two-phonon Raman peaks due zone-boundary modes.

\section{Theoretical approach}
\label{sec:theory}
We used the theoretical approach outlined in
Ref.~\cite{Subedi2014} to study the dynamics of the doubly-degenerate TA
modes of KTaO$_3$ at the $X$ point when its highest-frequency
TO mode at $X$ is externally pumped
through second-order Raman process. This density functional theory based
first-principles approach requires the calculation of the phonon
eigenvectors, which are then used to calculate the total energy
surface $V(\qhx,\qlx,\qlz)$ as a function of the high-frequency optical
and low-frequency acoustic modes.  The total energy surface is fit with a
polynomial to extract the phonon anharmonicities and phonon-phonon nonlinear
couplings (the full expression can be found in Appendix~\ref{sec:appendix1}), 
and these are used to construct the coupled equations of motion for the phonon
coordinates.  The coupling between the pumped mode and light is
extracted by calculating the total energy as a function of the $\qhx$ mode 
and electric field, an approach previously used by Cartella 
\etal~\cite{Cartella2018}. The coupled equations of motion are
solved numerically in the presence of a pump term for the $\qhx$ mode
to obtain the structural evolution of the material as a function of
time.

We used {\sc quantum espresso}~\cite{QE} (QE) for the computations of
the phonon frequencies and eigenvectors and the total energy surfaces
as a function of the phonon coordinates and electric field.
These were performed using ultrasoft pseudopotentials with the
valence orbitals $3s^{2}3p^{6}4s^{1}$ (K), $5s^{2}5p^{6}5d^{3}6s^{1}$
(Ta) and $2s^{2}2p^{4}$ (O) from the GBRV library~\cite{GBRV}. For the
exchange and correlation functional, we chose the PBEsol generalized
gradient approximation~\cite{PBEsol}. The plane-wave cutoffs for the
basis set and charge density expansions were set to 60 and 600 Ry,
respectively. As we are dealing with an
insulator with a gap, the electronic occupation was set to fixed.

The first step in our calculations was the relaxation of the unit cell, 
where we allowed the variation of both the lattice parameter and the
atomic positions. We let the relaxation process run until the difference
in the total energy between two steps of the self-consistent field (SCF) 
cycles was less than $10^{-10}$ Ry, the estimated error of the 
electronic density (which in our case is calculated as the electrostatic
self energy of the difference between the electronic densities at the
beginning and the end of each step of the calculation) was below $10^{-11}$ Ry,
and the components of 
the forces exerted on each atom were smaller than $10^{-6}$ Ry/Bohr.
We used a $12\times12\times12$ Monkhorst-Pack $k$-point grid for the
relaxation process. The lattice parameter obtained was $a=3.98784$ \AA, 
in good agreement with the experimental value
$a_{\textrm{exp}}=3.988$\AA~\cite{Verma2009}.

Once we had the relaxed unit cell, we used it for the computation of 
the phonon frequencies and eigenvectors at the Brillouin zone boundary
point $X$ $(0,\frac{1}{2},0)$, which was performed using density functional
perturbation theory~\cite{Savrasov1994} as implemented
in QE. The computation of the dynamical matrix requires a previous SCF
calculation which was performed using an $8\times8\times8$ Monkhorst-Pack
$k$-point grid. Then for the dynamical matrix calculation we set a threshold for
the self-consistent calculation of $10^{-18}$ Ry. The diagonalization of the
dynamical matrix was realized using the $\tt{dynmat}$ utility in QE,
thus obtaining the eigenvectors and frequencies of the different phonons.

For the computation of the phonon anharmonicities and phonon-phonon nonlinear
couplings, we used the calculated phonon eigenvectors to create modulated
structures as a function of the $\qhx$, $\qlx$, and $\qlz$ coordinates in
$1\times2\times1$ supercells that are required to simulate the phonons at the $X$
point, and then calculated the total energies of these structures. We sampled
values of the phonon coordinates ranging from $-3.0$ to 3.0
\AA$\sqrt{\textrm{u}}$. Steps of 0.025 and 0.1 \AA$\sqrt{\textrm{u}}$ were 
used for sampling the total-energy surfaces as a function of two and three
coordinates, respectively.
A convergence threshold of $10^{-10}$ Ry for the electronic density
in the SCF iterations and an $8\times 4\times 8$ Monkhorst-Pack $k$-point
grid was used in these calculations. To extract the anharmonicities and
nonlinear coupling constants, we fit the calculated total-energy surfaces
with polynomials having only the symmetry-allowed nonlinear terms using the 
{\sc glm}~\cite{GLM} package as implemented in {\sc julia}. The extracted
coefficients of the polynomials are given in Appendix~\ref{sec:appendix1}. 

We used the modern theory of polarization~\cite{Souza2002}
as implemented in QE to calculate the total energy of this material
as a function of the $\qhx$ coordinate and electric field $E$
and fit the resulting energy surface to the following expression:
\begin{equation}
    \label{eq:ph-E}
    \begin{split}
    H(\qhx,E) &= \frac{1}{2}\ohx^{2}\qhx^{2} + c_{4}\qhx^{4} + c_{6}\qhx^{6} + c_{8}\qhx^{8} \\
	& \quad + rE + sE^{2} + tE^{4} + \alpha \qhx^{2}E^{2}.    
    \end{split}    
\end{equation}
Here the frequency $\ohx$ and anharmonic coefficients $c_{i}$ of the
$\qhx$ mode are
those extracted from the previous total-energy calculations, and 
$s=-1.4829$ e\AA${}^{2}$/V, $t=-0.162$ e\AA${}^{4}$/V${}^{3}$ and $\alpha$
are the coefficients for the terms allowed by symmetry for
the electric field. The linear term for $E$ in $H(\qhx,E)$ with corresponding
coupling coefficient $r=-99.696$ e\AA\ occurs due to the use of periodic
boundary condition.  We sampled the electric field 
from $-36$ to 36 MV/cm with a step of 0.36 MV/cm and $\qhx$ from $-3.0$ to 3.0
\AA$\sqrt{\textrm{u}}$ with a step of 0.3 \AA$\sqrt{\textrm{u}}$. For
these calculations, we used an $8\times8\times8$ Monkhorst-Pack $k$-grid.
Like in the previous case, the {\sc glm} package was used to perform the fit.
The polynomial given in Eq.~\ref{eq:ph-E} fits the calculated total-energy
surface well, which is consistent with the fact that the form of the
coupling between the electric field and phonon at the
$X$ point is $\alpha\qhx^{2}E^{2}$ at the lowest order~\cite{Bartels2000}.
The fit gives a value for the coupling constant
$\alpha=0.074$ e/(V u). In order to check this method of
computing the light-phonon coupling, we also calculated the coupling of
the electric field to the highest frequency phonons of KTaO${}_3$ at
the $\Gamma$ point, obtaining Born effective mode charge of 
$Z^{*\rm{calc}}=1.03$ e/$\sqrt{\rm{u}}$, which is in good
agreement with the value of $Z^{*\rm{pert}}=1.07$ e/$\sqrt{\rm{u}}$
calculated
using density functional perturbation theory~\cite{Subedi2017}.
We note that the largest electric field used in the total-energy
calculations are more than an order of magnitude smaller than the values
that cause rectification of the $\qlz$ mode in the numerical solution of
the equations of motion discussed later. Larger values of the electric
field in total-energy calculations caused oscillations in the SCF
iterations.  This is a limitation of the currently available computational
method.  

The integration of the differential equations required for the 
solution of the equations of motion was carried out using the Strong
Stability Preserving method of Ruuth, an explicit Runge-Kutta order 3
propagator with 6 stages as implemented in the
{\sc DifferentialEquations}~\cite{DifferentialEquations} package from the
{\sc julia}
language. The time range for the propagation was from 0 to 8 ps, with a
time step of $8\times10^{-6}$ ps. The peak amplitude of the laser pulse 
was set to reach at 4 ps. For the initial conditions, we chose
$\qhx=\qlx=\qlz=0.1$ \AA$\sqrt{\textrm{u}}$, while their first
derivatives with respect to time were set to 0. In order to simulate the
thermal fluctuations of the phonons, we added a stochastic term in the
form of white noise to the equations of motion from the start of the
propagation until the pulse reaches its peak. Due to the presence of this
term, the solution obtained will depend on the particular string of random
values generated for each propagation. The criterion that we followed to 
determine the outcome of the propagation (in our case, whether or not there
is a rectification of the $\qlz$ mode) was to solve the equations multiple
times under the same pump amplitude and frequency conditions, but with a
different seed for the random number generator for each run. Then we pick
the most probable solution among those obtained, i. e., the one that occurs
the most number of times in at the end of our propagations.
The Fourier transform of the solutions was obtained using
the {\sc fftw}~\cite{FFTW} package as implemented in {\sc julia}.

\section{Results and Discussion}
\label{sec:results}

\begin{figure}
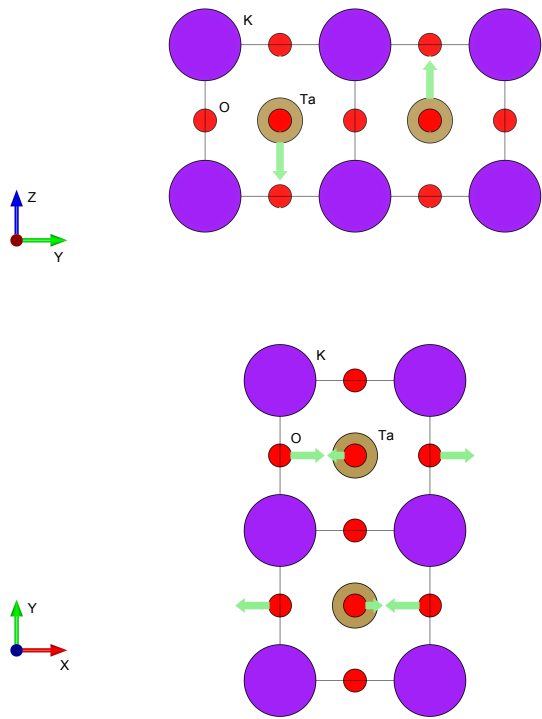

	\centering
	\begin{minipage}{1.3\linewidth}
		\includegraphics[width=\linewidth]{YLZ.pdf}
	\end{minipage}
	\begin{minipage}{1.3\linewidth}
		\includegraphics[width=\linewidth]{YHX.pdf}
	\end{minipage}
        \caption{Schematic representations of the phonon modes of KTaO$_3$ at the
        $X$ $(0,1/2,0)$ point considered in the present work.
        (Top) The TA mode component $\qlz$ that has atomic movements polarized
        along the $z$ direction.  The other degenerate component of this mode $\qlx$ has the same atomic movements but are directed along the $x$ axis.
        The TA mode is the lowest-frequency mode at $X$ in KTaO$_3$.
        (Bottom) The highest-frequency TO mode $\qhx$ that has atomic 
        movements polarized along the $x$ direction.}
\label{fig:phononY} 
\end{figure}

The TA and TO modes of KTaO$_3$ at $X$ are doubly degenerate.  The TA 
mode is the lowest-frequency phonon at $X$, whereas there are four TO 
phonon branches in this material.
Figs.~\ref{fig:phononY}(top) and (bottom) show the atomic displacements 
corresponding to the $\qlz$ and $\qhx$ components of the TA and 
highest-frequency TO modes, respectively.
The calculated frequencies of these modes are $\Omega_{\textrm{LZ}}$ = 61
cm$^{-1}$ and $\Omega_{\textrm{HX}}$ = 509 cm$^{-1}$, respectively.
These are in good accord with the values inferred from the Raman
experiments of Nilsen and Skinner, where these modes manifest as peaks
at 123 and 1095 cm$^{-1}$ corresponding to the doubling of the
respective phonon frequencies due to second-order Raman processes
\cite{Nilsen1967}.  Both these modes belong to the irreducible 
representation $X^{+}_{5}$ of the cubic structure with the space group 
$Pm\overline{3}m$. The $\qlz$ mode involves displacement of the Ta ions
against the O octahedra along the $z$ direction.  The $\qhx$ mode causes
one set of planar O ions to move against the Ta ions in the $x$
direction, while another set of planar O ions remain stationary.  This mode
also displaces the apical O ions along the $x$ direction against the
movement of the planar O ions.  Since these modes have the wavevector
$(0,\frac{1}{2},0)$, the atomic displacements within the adjacent unit
cells are out-of-phase along the $y$ direction, thus breaking the
translation symmetry.  The distorted structure has the orthorhombic
space group $Pmma$.

\begin{figure}[htp]
	\includegraphics[scale=0.7]{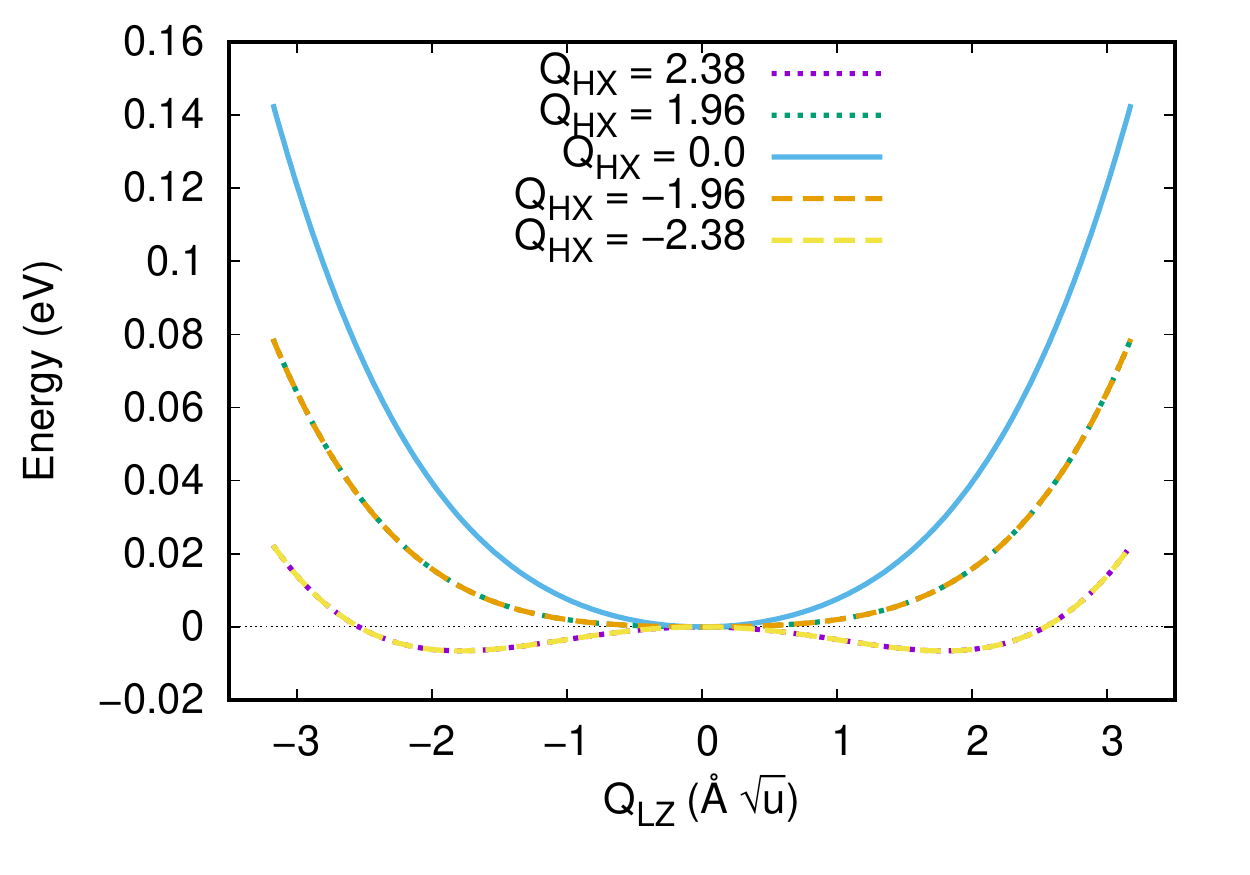}
	\caption{\label{fig:double_well} Total energy as a function
          of the $\qlz$ phonon coordinate for different
          values of the $\qhx$ phonon coordinate. For visual clarity, 
          the zero energy point 
          has been chosen so that the curves coincide at $\qlz=0$.}
\end{figure}

We calculated the total energy as a function of the $\qhx$ and
$\qlz$ coordinates, and Fig.~\ref{fig:double_well} shows
five energy curves from this energy surface $V(\qhx,\qlx=0,\qlz)$.  We
can see that the total energy as a function of the $\qlz$
coordinate for a fixed value of the $\qhx$ coordinate is
symmetric upon the transformation $\qlz \rightarrow
-\qlz$.  The $\qhx$ and $-\qhx$
energy curves also overlap with each other.  This implies that the
energy surface is an even function of both $\qlz$ and
$\qhx$, and these coordinates occur only with even powers
in the polynomial fit of the energy surface.  This is consistent with
the symmetry requirement that the coupling terms occur with even
powers of the coordinates when they are orthogonal to each other.

The energy curve of the $\qlz$ coordinate softens when the
$\qhx$ coordinate has a finite value, and it develops a
double-well shape at large values of the $\qhx$ coordinate.
This is reflected in the negative sign of the coefficients in the
nonlinear coupling terms $g_1 \qhx^2 \qlz^2$, $g_2 \qhx^4 \qlz^2$, and
$g_3 \qhx^2 \qlz^4$ in the fit of $V(\qhx,\qlx=0,\qlz)$ (see Appendix 
\ref{sec:appendix1}).  The total
force experienced along the $\qlz$ coordinate is given by $-\partial
V/\partial Q_{LZ}$, and the effect of the nonlinear terms is to
renormalize its frequency as  $\olz^2 \rightarrow \olz^2 (1 + 2 g_1
\qhx^2 + 2 g_2 \qhx^4 + 4 g_3 \qhx^2 \qlz^2 + \cdots)$.  Since the phonon
coordinates $\qhx$ and $\qlz$ appear with even powers in this
expression, their contribution to the renormalization will not be
averaged out over time.  As a result, the low-frequency mode $\qlz$ softens
when the high-frequency mode $\qhx$ is oscillating with a finite amplitude.

\begin{figure}[htp]
	\includegraphics[scale=0.7]{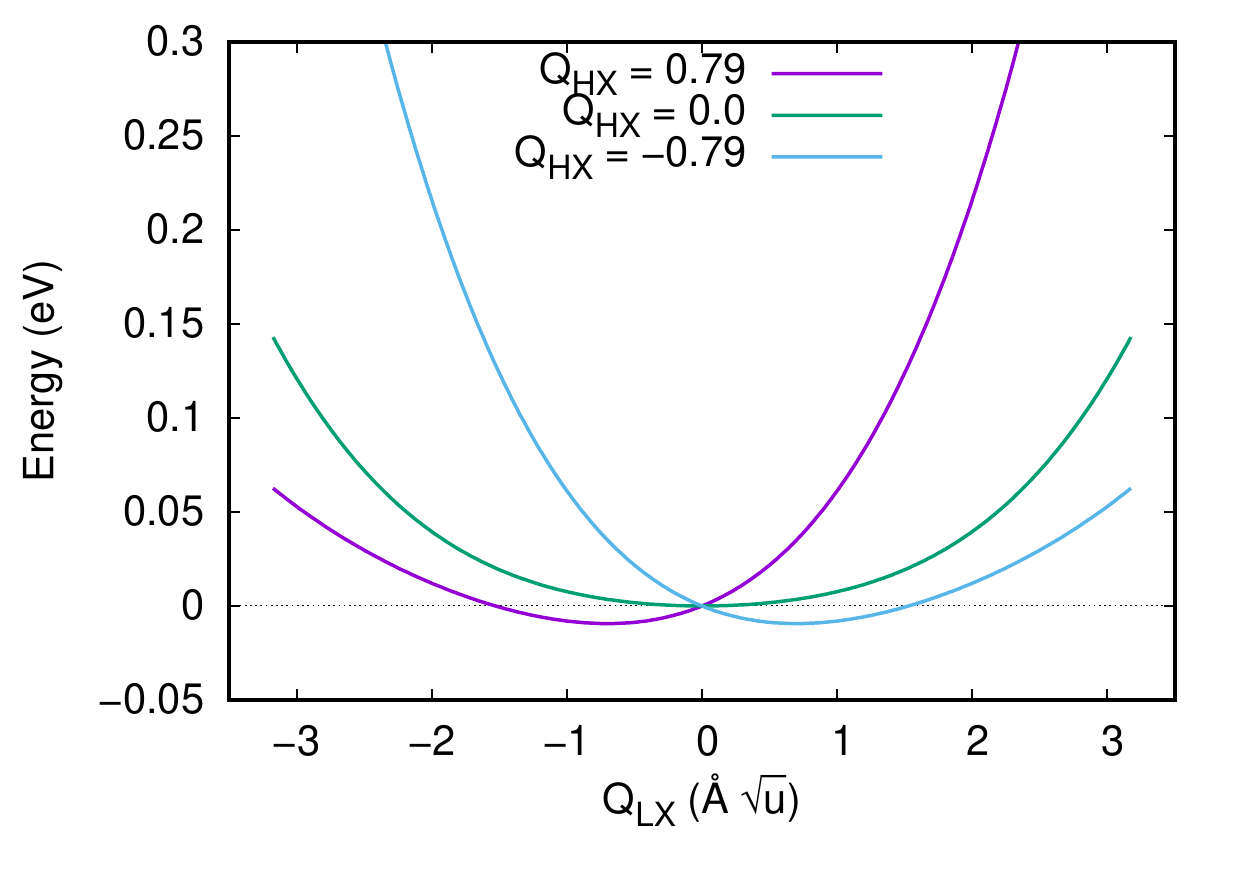}
	\caption{\label{fig:displaced_minimum} Total energy as a function
          of the $\qlx$ phonon coordinate  for different
          values of the $\qhx$ phonon coordinate. For visual clarity, 
          the zero energy point has been chosen so
          that the curves coincide at $\qlx = 0$.}
\end{figure}

We also investigated
the dynamics along the $\qlx$ component of the TA mode that has atomic 
displacements parallel to that of the high-frequency $\qhx$ mode.
The energy curves of the $\qlx$ coordinate for several values of the
$\qhx$ coordinate extracted from the calculated total-energy surface 
$V(\qhx, \qlx, \qlz = 0)$ is shown in Fig.~\ref{fig:displaced_minimum}.
In this case we can see that the minimum of the $\qlx$ coordinate
shifts when the $\qhx$ coordinate has a finite value, and the direction of
this displacement depends on the sign of $\qhx$.  The curves with
the same magnitude of $\qhx$ but opposite sign are mirror images of
each other, with the mirror plane located at $\qlx = 0$.  
In addition, the energy curves of the $\qlx$ mode noticeably harden as 
the magnitude of the $\qhx$ coordinate is increased, which contrasts 
with the softening exhibited by $\qlz$ mode.  
This implies that the energy surface includes terms of the form
$\qhx^i \qlx^j$ with both even and odd powers, subject to the
condition $i+j=2n$, where $n$ is an integer. Once again, this is in
accord with the symmetry requirements for two modes with the same
irreducible representation and parallel polarization.

We constructed the coupled equations of motion for the $\qhx$, $\qlx$, 
and $\qlz$ coordinates using the calculated total-energy surfaces as
the potential energy.  These equations read
\begin{align}
  \label{eq:eq_motion}
  \ddot{Q}_{\textrm{HX}} + \gamma_{\textrm{HX}}\dot{Q}_{\textrm{HX}} + \ohx^2 \qhx &= -\frac{\partial V^{{\textrm{nh}}}(\qhx, \qlx, \qlz)}{\partial \qhx} \nonumber\\
   & \quad + F(t), \nonumber \\
  \ddot{Q}_{\textrm{LX}} + \gamma_{\textrm{LX}}\dot{Q}_{\textrm{LX}} + \olx^2 \qlx &= -\frac{\partial V^{{\textrm{nh}}}(\qhx, \qlx, \qlx)}{\partial \qlx},\nonumber\\
  \ddot{Q}_{\textrm{LZ}} + \gamma_{\textrm{LZ}}\dot{Q}_{\textrm{LZ}} + \olz^2 \qlz &= -\frac{\partial V^{{\textrm{nh}}}(\qhx, \qlx,
   \qlz)}{\partial \qlz}.    
\end{align}
Here $V^{{\textrm{nh}}}(\qhx,\qlx,\qlz)$ is the nonharmonic part of
the polynomial fit to the calculated total-energy surfaces as a
function of the three coordinates and $\gamma_{i}$'s are the damping
coefficients of the corresponding normal modes, which we set to
10\% of the value of their corresponding natural frequency. 
The full polynomial expression of $V^{\textrm{nh}}$ with terms up to 
the eight order that was used for fitting the calculated total-energy 
surfaces is given in Appendix~\ref{sec:appendix1}.
$F(t)$ is the external force experienced by the $\qhx$ coordinate due
to the pump pulse.  This was taken into account by considering the
force on $\qhx$ due to an electric field, which is given by 
\begin{equation}
    \begin{split}
        F & =  - \frac{\partial H(\qhx,E)}{\partial \qhx} \\
          & = - 2 \alpha \qhx E^2. 
    \end{split}
\end{equation}
We studied the dynamics using Gaussian-enveloped single-frequency
pulses
\begin{align}
\label{eq:pumpsf}
E_{\rm sf}(t) & = E_{0}\sin(\omega t)e^{-t^{2}/2(\sigma/2\sqrt{2\log
    2})^{2}}.
\end{align}
Here, $E_{0}$ is the amplitude of the pulse and
$\omega$ its frequency.  The pulse has a
Gaussian envelope with full-width at half maximum of $\sigma$.

\begin{figure*}
        \includegraphics[width=\textwidth]{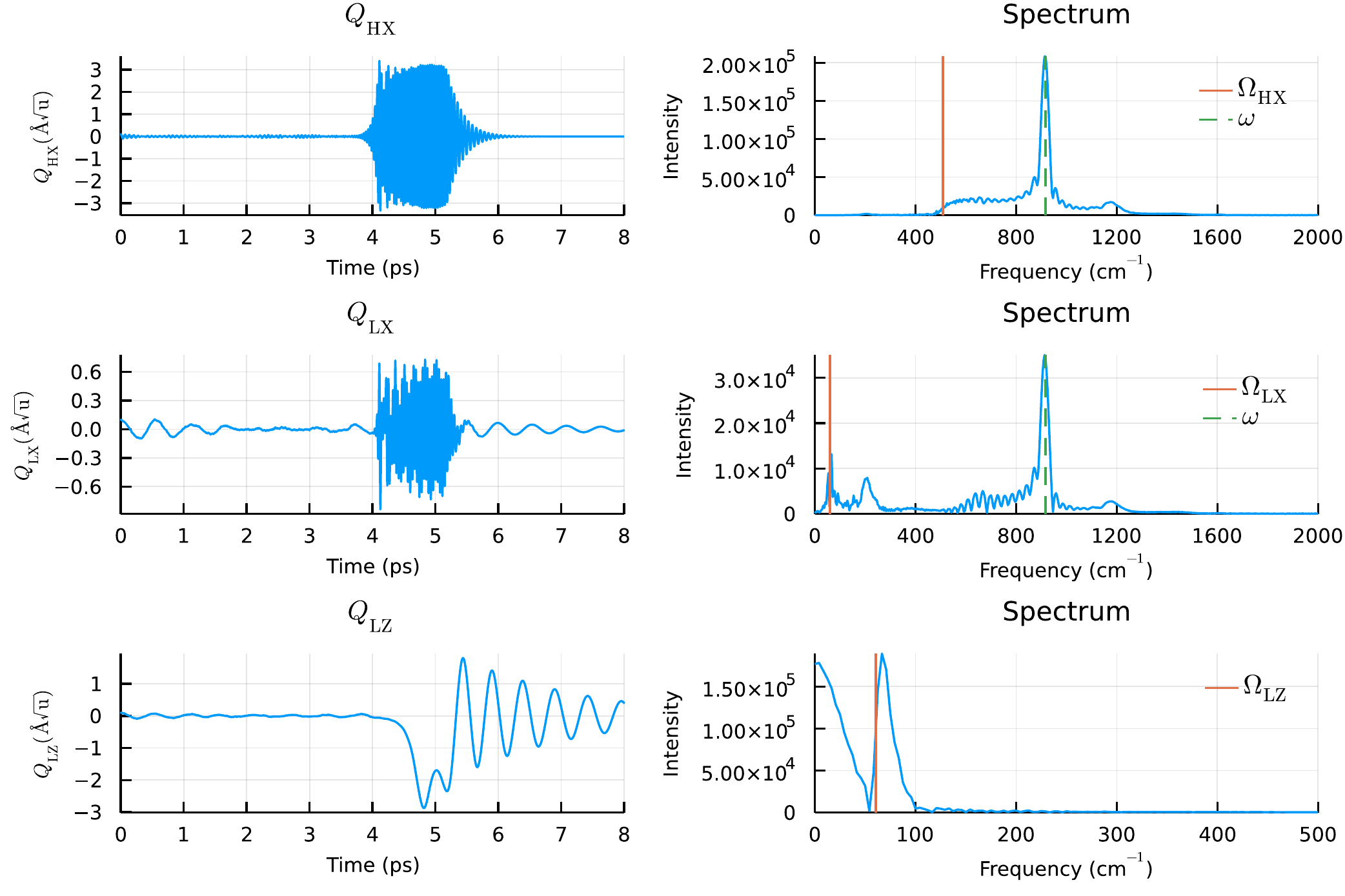}
        \caption{\label{fig:y_sm1} (Left) Dynamics of the (top) TO $\qhx$, 
        (middle) TA $\qlx$, and (bottom) TA $\qlz$ phonon coordinates at the $X$
        point for a single-frequency pump pulse with amplitude $E_{0}=465$
        MV/cm and frequency $\omega=1.8\ohx$.
        (Right) Fourier transform of the time evolution of the respective
        coordinates. The solid vertical lines mark the natural frequencies
        of each mode, while the dashed one indicates the
        frequency of the pump pulse.}
\end{figure*}

The coupled equations of motion for the $\qhx$, $\qlx$, and $\qlz$
coordinates given in Eq.~\ref{eq:eq_motion} were solved for different
values of pump amplitude $E_0$ and frequency $\omega$.  For small
values of the pump amplitude $E_0$, the energy transferred to $\qhx$
by the external pulse is small. This mode then oscillates at its
natural frequency $\ohx$ without getting amplified regardless of the
frequency of the pump pulse and decays at a rate determined by
$\gamma_{\textrm{HX}}$.  As a result, the force imparted on the $\qlz$
and $\qlx$
coordinates due to the oscillation of $\qhx$ is also small, and $\qlz$
and $\qlx$ also exhibit decaying oscillations about their 
natural frequency
$\olx=\olz$.  For very large values of pump amplitude $E_0$, all three 
modes diverge, which describes the breakdown of the
material at very high electric field of the pump. In between these two
limiting behaviors, we searched for a range of pump frequency and
amplitude that causes the $\qlz$ mode to oscillate at a displaced
position.

We find that $\omega = 1.8 \ohx$ is the lowest pump frequency that
leads to a rectification of the $\qlz$ coordinate, which occurs for a
pump amplitude of 465 MV/cm. The solutions of the equations of motion
for the $\qhx$, $\qlx$ and $\qlz$ coordinates for these values of pump
frequency and amplitude are shown in Fig.~\ref{fig:y_sm1}. As one can
see, the low-frequency $\qlz$ coordinate oscillates at a displaced
position while the externally-pumped $\qhx$ coordinate is oscillating
with a large amplitude. This implies that the translation symmetry of
the lattice is broken because the $\qlz$ coordinate has a non-zero
average value within this duration. When the $\qhx$ mode decays after
the diminution of the pump pulse, the $\qlz$ coordinate goes back to
oscillating about the equilibrium position with a decaying amplitude.
In the Fourier transform of $\qlz(t)$, the displaced motion appears as
a large intensity around zero frequency, while the amplified
oscillations after the pump appear as a peak near the original
frequency $\olz$.

Fig.~\ref{fig:y_sm1} also shows that the externally-pumped phonon mode
$\qhx$ is highly amplified and oscillates with an amplitude of $\sim$3
\AA$\sqrt{\textrm{u}}$. Its Fourier transform shows a resonance
peak at the frequency of the pump pulse in this regime, but frequency
components between $\ohx$ to $\sim$2.5$\ohx$ also show significant
contribution. This reflects the parametrically-driven nature of the
equation of motion of the $\qhx$ mode because the external force
$F(t) = -2\alpha \qhx E(t)^2$ due to the pump pulse is linear in
$\qhx$. As a result, the frequency of the driven $\qhx$ mode varies
with time and acquires components that are not resonant with respect
to the harmonic frequency of the mode or the pump frequency.  
The other TA coordinate $\qlx$ with atomic motions parallel to
the $\qhx$ mode, whose dynamics is also shown in the figure, is 
moderately amplified while the $\qhx$ mode is making large-amplitude
oscillations.  Fourier transform of the time evolution of this mode 
shows a large peak at the pump frequency.  This high-frequency oscillation
of the TA $\qlx$ mode reflects the large $\qhx^2\qlx^2$ nonlinearity.  

At a pump frequency of $\omega = 1.8 \ohx$,
the $\qlz$ coordinate makes
only a single cycle of oscillation at a displaced position during
the pump pulse. This
indicates that the effective double-well potential experienced by this
mode is shallow. Indeed, we find that the range of pump amplitude that 
causes the rectification of the $\qlz$ mode is relatively narrow for
this value of pump frequency.  The $\qlz$ coordinate again oscillates 
about the equilibrium position as the pump amplitude is increased 
above 525 MV/cm.  However, the amplitude of oscillations remain larger
than 3 \AA$\sqrt{\textrm{u}}$, indicating that $\qlz$ mode oscillates
across the minima of the double-well potential at these higher values of
pump amplitude.

\begin{figure}[htp]
	\includegraphics[width=\columnwidth]{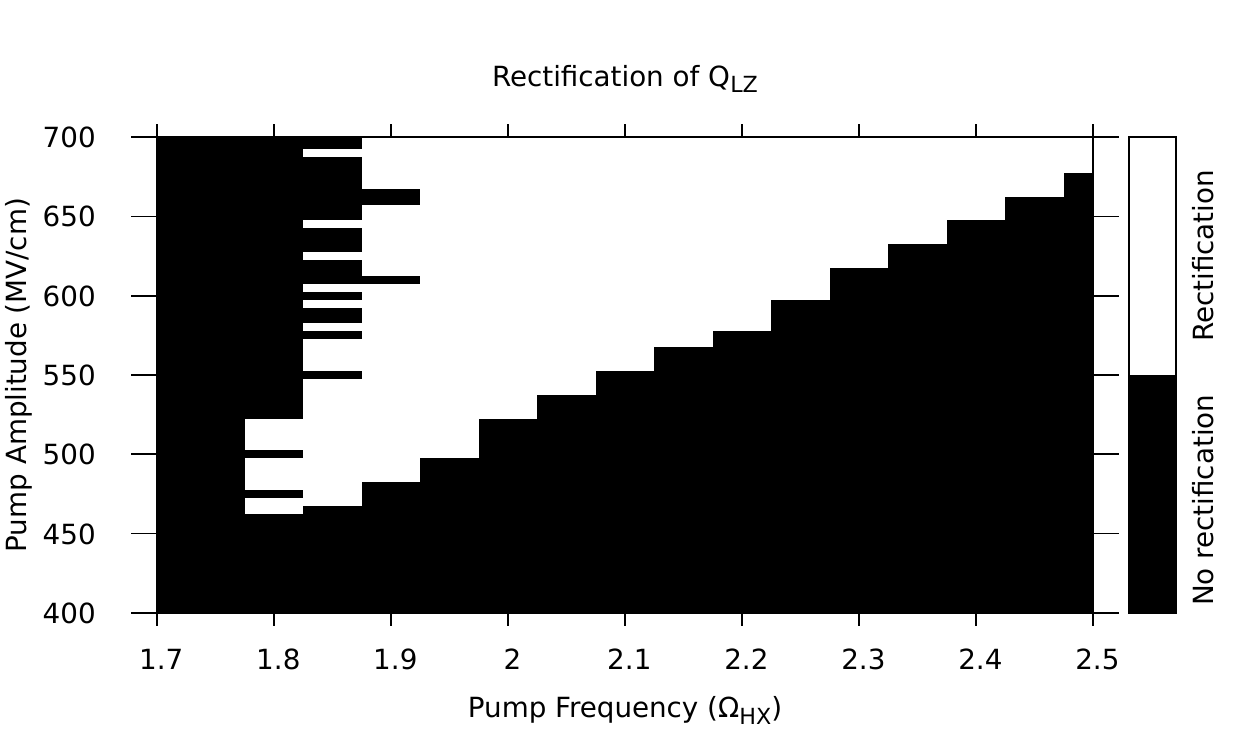}
	\caption{\label{fig:rectification} Values for the amplitude and frequency
	of the single-frequency pulse used to pump the $\qhx$ phonon coordinate 
	that induce rectification of the $\qlz$ coordinate.}
\end{figure}

Fig.~\ref{fig:rectification} shows the ranges of pump amplitudes that
rectify the $\qlz$ coordinate when the $\qhx$ coordinate is pumped at
frequencies between 1.7$\ohx$ and 2.5$\ohx$ and amplitudes between 400
and 700 MV/cm. 
We actually solved the equations of motion of the phonon coordinates 
for pump frequencies up to 3.0$\ohx$ and amplitudes up to 1500 MV/cm.
As already mentioned, the equations of motion include a white noise term
to simulate the thermal fluctuations of the phonons. Up to a pump amplitude
of 700 MV/cm, the presence (or absence) of 
rectification of the $\qlz$ mode is independent of the noise term with the 
exception of the values of pump amplitude and frequency near the border between
rectification and no rectification, where both outcomes appear in the
solutions of the equations of motion.
Larger pump values cause the appearance of divergences in the
solution that pervade the entirety of the range of frequencies studied, and we
enter a new regime of the dynamics of the system that we will analyze below.
For this reason Fig.~\ref{fig:rectification} is limited to the values of pump
amplitude and frequency that induce rectification without possible breakdown
of the material.

We can see that the value of the smallest pump amplitude that rectifies 
the $\qlz$ coordinate increases with the pump frequency. It is 465 MV/cm for 
$\omega = 1.8\ohx$ and increases to 675 MV/cm for $\omega = 2.5\ohx$.
This increasing dependence derives from the fact that a larger pump
amplitude is required to resonantly excite the $\qhx$ coordinate at
higher pump frequencies.
The largest pump amplitude that rectifies the
$\qlz$ coordinate increases steeply as a function of the pump
frequency. It is 520 MV/cm for $\omega = 1.8 \ohx$ and increases
to a value of more than 700 MV/cm for $\omega = 1.9\ohx$, where 
the solutions become dependent on noise as discussed in the previous
paragraph. In fact the rectified solutions for the $\qlz$ coordinate
appear at pump amplitudes up to 840, 1410, and 1490 MV/cm for $\omega
= 1.9\ohx$, $2.1\ohx$, and $2.5\ohx$, respectively. At higher pump
frequencies, the largest pump amplitude that gives a rectified solution 
flatlines at 1490 MV/cm up till the largest pump frequency of 3.0$\ohx$ 
that we tested.
On the other hand, the lowest pump amplitude that rectifies $\qlz$ 
keeps slowly increasing to a value of 840 MV/cm for $\omega = 3.0 \ohx$.
Therefore, the window of pump amplitude that rectifies the $\qlz$ mode
is narrow when rectification starts occurring at $\omega = 1.8 \ohx$,
broadens up to $\omega = 2.5 \ohx$, and starts narrowing  again as
the pump frequency is further increased.

\begin{figure*}[htp]
	\includegraphics[width=\textwidth]{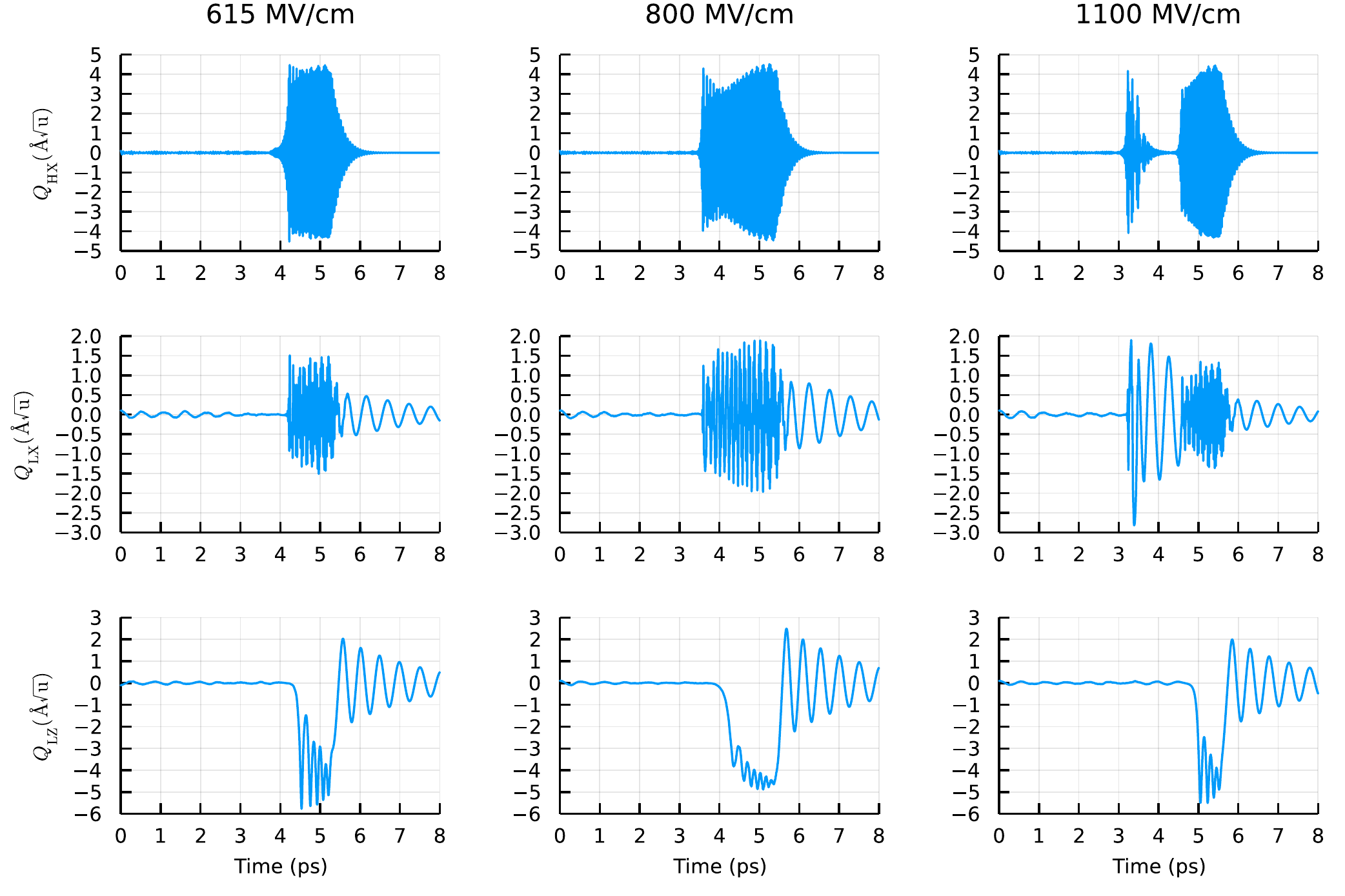}
	\caption{\label{fig:comparison} Dynamics of the $\qhx$, $\qlx$ and
	$\qlz$ phonon coordinates for pump pulses with frequency
	$\omega=2.3\ohx$ and amplitudes of 615, 800 and 1100 MV/cm.}
\end{figure*}

We now illustrate the light-induced dynamics for the case of a pump frequency 
that exhibits a large window of rectification of the $\qlz$ coordinate as a
function of the pump amplitude. 
The three columns of Fig.~\ref{fig:comparison} show the solutions of the coupled
equations of motion of the three phonon coordinates at
a pump frequency of 2.3$\ohx$ for pump amplitudes of 615, 800, and
1100 MV/cm.  At 615 MV/cm, which is the lower threshold of the rectification
window for this pump frequency, the $\qlz$ mode exhibits six cycles of
oscillations while it is rectified [bottom panel in 
Fig.~\ref{fig:comparison}(left)].  This indicates that the larger
value of the pump amplitude required for rectification at a higher
pump frequency makes the effective double-well potential deeper, which
increases the frequency of the $\qlz$ mode when it is rectified.
Furthermore, the oscillations of the $\qlz$ mode occur about $\sim$4
\AA$\sqrt{\textrm{u}}$, indicating that the minima of the effective
double-well potential gets further away from the equilibrium value of
zero for a larger value of the pump amplitude.

Fig.~\ref{fig:comparison} (middle) shows the dynamics at an increased
pump amplitude of 800
MV/cm while keeping the pump frequency fixed at 2.3$\ohx$.  The $\qlz$
mode now makes eight cycles while displaced from the equilibrium
position.  This happens not due to an increase in the frequency of the
oscillations in the rectified regime but because the mode gets
rectified for a longer duration.  There is only
a marginal change in the position about which this mode oscillates
while it is rectified.  Furthermore, the maximum amplitude of the
oscillations of the $\qlz$ mode gets reduced.  Interestingly, the
amplitude of the oscillation of the pumped $\qhx$ mode also does not
increase as the pump amplitude is increased from 615 to 800 MV/cm.
Instead, its amplitude as a function of time exhibits a small dip 
before increasing again by a similar amount.  The
amplification of the $\qhx$ also occurs for a longer duration.
Therefore, the additional pump energy causes rectification and
amplification for a longer duration rather than displacing the $\qlz$
mode to a larger distance or increasing the amplification of the
$\qhx$ mode.  The additional pump energy also flows to the $\qlx$ mode, 
whose amplified oscillations last for a longer duration as well.

Fig.~\ref{fig:comparison} (right) shows the dynamics when the pump
amplitude is further
increased to 1100 MV/cm while keeping the pump frequency at 2.3$\ohx$.
The amplified oscillations of the pumped $\qhx$ mode now splits into two
different packets that are separated by a region where the mode is
unamplified.  The $\qlz$ mode gets rectified, but only during the
amplified oscillations of the $\qhx$ mode in the packet of the later
time delay.  As	a result, the $\qlz$ mode makes	only four cycles of
oscillations at	a displaced position.  Neither the amount of
displacement from the equilibrium position nor the amplitude of
oscillations of	the $\qlz$ mode	increase in the	rectified regime at
this increased value of the pump amplitude.	The amplitude of the
$\qhx$ mode also does not increase, while this mode in fact now makes
amplified oscillations for a shorter duration. 	However, the $\qlx$
mode now oscillates with a much larger amplitude during the initial
part of the pump pulse, and this accounts
for the	additional energy pumped into the system.

At higher pump frequencies, we find the same trend shown in
Fig.~\ref{fig:comparison} as the pump amplitude is increased.
Interestingly, the
amplitude of the oscillations of the $\qhx$ mode and the displacement
of the $\qlz$ mode do increase as the pump frequency is increased, but
they vary little as the pump amplitude is increased while keeping the
pump frequency fixed.

The very high values of the pump amplitude that we find necessary
to break the translation symmetry of KTaO$_3$ are not achievable using
currently available laser sources in the midinfrared regime.  Thus, our work
provides motivation for the development of intense midinfrared
lasers.  Large electric fields using available midinfrared sources can
also be achieved if the sample can be grown inside metallic cavities,
and this study should further stimulate the ongoing work to perfect
advanced thin-film growth techniques.
Furthermore, the high value of the electric field may damage the
sample even though the excitation is done for a relatively short
duration at frequencies much lower than the band gap of the material.
Nevertheless, our study does show that translation symmetry breaking
by externally pumping a zone-boundary phonon mode of a material is
possible in principle via the mechanism of nonlinear phononics.  
The physical parameter that limits the
efficiency of this phenomenon is the smallness of the coupling between
light and two-phonon excitation of the zone-boundary mode.  This work
motivates the search for a material that exhibits stronger
second-order Raman scattering of the zone-boundary phonon than that
found in KTaO$_3$.

\section{Summary and Conclusions}
\label{sec:conclusion}

In summary, we have investigated the possibility of light-induced
translation symmetry breaking via nonlinear phononics in  KTaO$_3$
by pumping its zone-boundary TO phonon mode.
This work was motivated 
by the previously reported experimental observation of Brillouin 
zone boundary phonon modes in the Raman spectra of this material
due to second-order Raman processes.  We calculated the total energy
of this material as a function of the highest-frequency TO mode 
$\qhx$ and degenerate components of the TA mode $\qlx$ and $\qlz$ 
from first principles to obtain phonon anharmonicities and phonon-phonon 
nonlinear couplings.  We find that the energy curve of the $\qlz$
mode softens and develops a double-well shape as the value of the
$\qhx$ coordinate is increased, indicating that $\qlz$ mode
becomes unstable when the $\qhx$ mode is pumped with sufficiently
intense laser pulses.  The coupling between the $\qhx$ mode and light 
was similarly obtained from first principles by 
calculating the total energy of this material as a function of the
$\qhx$ coordinate and electric field.  These were then used to construct
coupled equations of motion of the phonon coordinates in the presence
of a Gaussian-enveloped single-frequency pump pulse term on the $\qhx$ mode.

We solved the coupled equations of motion for a range of pump frequency 
and amplitude.  We find that $1.8\ohx$ is the smallest pump frequency
for which the $\qlz$ oscillates at a displaced position, and this 
occurs for a pump amplitude range of 465--520 MV/cm.
Since the $\qlz$ coordinate has a nonzero time-average when it is
rectified, this implies that the translation symmetry of this material
is broken for this duration. As the pump frequency is increased, the 
magnitude of the smallest pump amplitude that rectifies the $\qlz$ mode
also increases.  These values of pump intensity are at least an order 
of magnitude larger than that can be produced by currently available 
midinfrared laser sources.  Moreover, the high value of electric field
may cause dielectric breakdown of the sample even for a pump pulse
of short duration at a frequency much smaller than the band gap of the
material. 
Nonetheless this study shows that light can in principle be used 
to break the translation symmetry of a material by pumping 
a phonon mode at the Brillouin zone boundary, opening the door to
a new form of materials control via nonlinear phononics. 

\begin{acknowledgments}
This work was supported by the Agence Nationale de la Recherche
under grant ANR-19-CE30-0004 ELECTROPHONE and GENCI-TGCC under grant
A0110913028.
\end{acknowledgments}

\cleardoublepage

\appendix
\section{Expression for the Total Energy Surface}
\label{sec:appendix1}

The calculated total-energy surface $V(\qhx, \qlx, \qlz)$
was fit with the expression
\begin{equation}
    \label{eq:V}
    \begin{split}
        V & = \frac{1}{2}\olx^{2}\qlx^{2} + \frac{1}{2}\olz^{2}\qlz^{2} + \frac{1}{2}\ohx^{2}\qhx^{2} \\
          & \quad + V^{{\textrm{nh}}},
    \end{split}
\end{equation}
where the nonharmonic part $ V^{{\textrm{nh}}}(\qhx, \qlx, \qlz)$ is given by
\begin{equation}
\label{eq:Vnh}
\begin{split} 
        V^{{\textrm{nh}}} & = a_{4}\qlx^{4} + a_{6}\qlx^{6} + a_{8}\qlx^{8} \\
        & \quad + b_{4}\qlz^{4} + b_{6}\qlz^{6} + b_{8}\qlz^{8} \\
        & \quad + c_{4}\qhx^{4} + c_{6}\qhx^{6} + c_{8}\qhx^{8} \\
        & \quad + e_{1}\qlx^{2}\qlz^{2} + e_{2}\qlx^{4}\qlz^{2} + e_{3}\qlx^{2}\qlz^{4} \\
        & \quad + e_{4}\qlx^{6}\qlz^{2} + e_{5}\qlx^{4}\qlz^{4} + e_{6}\qlx^{2}\qlz^{6}\\
        & \quad + f_{0}\qhx\qlx + f_{1}\qhx^{3}\qlx + f_{2}\qhx^{2}\qlx^{2} \\
        & \quad + f_{3}\qhx\qlx^{3} + f_{4}\qhx^{5}\qlx + f_{5}\qhx^{4}\qlx^{2} \\
        & \quad + f_{6}\qhx^{3}\qlx^{3} + f_{7}\qhx^{2}\qlx^{4} + f_{8}\qhx\qlx^{5} \\
        & \quad + f_{9}\qhx^{7}\qlx + f_{10}\qhx^{6}\qlx^{2} + f_{11}\qhx^{5}\qlx^{3} \\
        & \quad + f_{12}\qhx^{4}\qlx^{4} + f_{13}\qhx^{3}\qlx^{5} + f_{14}\qhx^{2}\qlx^{6} \\
        & \quad + f_{15}\qhx\qlx^{7}\\
        & \quad + g_{1}\qhx^{2}\qlz^{2} + g_{2}\qhx^{4}\qlz^{2} + g_{3}\qhx^{2}\qlz^{4}\\
        & \quad + g_{4}\qhx^{6}\qlz^{2} + g_{5}\qhx^{4}\qlz^{4} + g_{6}\qhx^{2}\qlz^{6}\\
        & \quad + j_{1}\qhx\qlx\qlz^{2} + j_{2}\qhx^{3}\qlx\qlz^{2} \\
        & \quad + j_{3}\qhx^{2}\qlx^{2}\qlz^{2} + j_{4}\qhx\qlx^{3}\qlz^{2} \\
        & \quad + j_{5}\qhx\qlx\qlz^{4} + j_{6}\qhx^{5}\qlx\qlz^{2} \\
        & \quad + j_{7}\qhx^{4}\qlx^{2}\qlz^{2} + j_{8}\qhx^{3}\qlx^{3}\qlz^{2} \\
        & \quad + j_{9}\qhx^{2}\qlx^{4}\qlz^{2} + j_{10}\qhx\qlx^{5}\qlz^{2} \\
        & \quad + j_{11}\qhx^{3}\qlx\qlz^{4} + j_{12}\qhx^{2}\qlx^{2}\qlz^{4} \\
        & \quad + j_{13}\qhx\qlx^{3}\qlz^{4} + j_{14}\qhx\qlx\qlz^{6}.
\end{split}
\end{equation}

The terms appearing in this expression are those allowed
by the symmetry. We found that terms up
to the eighth order, with the coefficients smaller than $10^{-7}$
neglected, suffice to describe calculated total-energy surface.
The values of the coefficients appear in Table~\ref{tab:coefsY}.

\begin{table*}[htp]
\caption{\label{tab:coefsY}
  The coefficients of the harmonic, anharmonic, and nonlinear coupling terms
  of the polynomial used to fit the calculated total-energy surface $V(\qhx, \qlx, \qlz)$
  of KTaO$_3$ as a function of the three $X$-point phonon coordinates considered in this study. The units are
  $\textrm{eV}\left(\frac{\text{\AA}}{\sqrt{\textrm{u}}}\right)^{i+j+k}$, where $i$, $j$ and $k$ are
  the exponents of the phonon coordinates.}
\begin{ruledtabular}
\begin{tabular}{CCC|CCC}
  \text{Coefficient}  &\text{Order} & \text{Value} & \text{Coefficient}  &\text{Order} & \text{Value}\\
\hline
\olx^{2} & \qlx^{2} & 0.013636 & f_{9}& \qhx^{7}\qlx & 2.92\times10^{-5}\\
\olz^{2} & \qlz^{2} & 0.013636 & f_{10}& \qhx^{6}\qlx^{2} & 3.84\times10^{-5}\\
\ohx^{2} & \qhx^{2} & 0.955643 & f_{11}& \qhx^{5}\qlx^{3} & 2.65\times10^{-5}\\

a_{4}& \qlx^{4} & 7.95\times10^{-4} & f_{12}& \qhx^{4}\qlx^{4} & 1.4\times10^{-5}\\
a_{6}& \qlx^{6} & -7.75\times10^{-6} & f_{13}& \qhx^{3}\qlx^{5} & 3.59\times10^{-6}\\
a_{8}& \qlx^{8} & 1.31\times10^{-7} & f_{14}& \qhx^{2}\qlx^{6} & 6.7\times10^{-7}\\

b_{4}& \qlz^{4} & 7.95\times10^{-4} & g_{1}& \qhx^{2}\qlz^{2} & -4.789\times10^{-4}\\
b_{6}& \qlz^{6} & -7.75\times10^{-6} & g_{2}& \qhx^{4}\qlz^{2} & -2.458\times10^{-4}\\
b_{8}& \qlz^{8} & 1.31\times10^{-7} & g_{3}& \qhx^{2}\qlz^{4} & -3.07\times10^{-5}\\

c_{4}& \qhx^{4} & 0.044353 & g_{4}& \qhx^{6}\qlz^{2} & -2.01\times10^{-6}\\
c_{6}& \qhx^{6} & 2.649\times10^{-4} & g_{5}& \qhx^{4}\qlz^{4} & 1.82\times10^{-6}\\
c_{8}& \qhx^{8} & 1.58\times10^{-5} & g_{6}& \qhx^{2}\qlz^{6} & 3.18\times10^{-7}\\

e_{1}& \qlx^{2}\qlz^{2} & 2.796\times10^{-4} & j_{1}& \qhx\qlx\qlz^{2} & 0.001085\\
e_{2}& \qlx^{4}\qlz^{2} & -1.14\times10^{-5} & j_{2}& \qhx^{3}\qlx\qlz^{2} & -2.49\times10^{-4}\\
e_{3}& \qlx^{2}\qlz^{4} & -1.14\times10^{-5} & j_{3}& \qhx^{2}\qlx^{2}\qlz^{2} & -1.233\times10^{-4}\\
e_{4}& \qlx^{6}\qlz^{2} & 1.96\times10^{-7} & j_{4}& \qhx\qlx^{3}\qlz^{2} & -3.4\times10^{-5}\\
e_{5}& \qlx^{4}\qlz^{4} & 2.35\times10^{-7} & j_{5}& \qhx\qlx\qlz^{4} & -8.7\times10^{-5}\\
e_{6}& \qlx^{2}\qlz^{6} & 1.96\times10^{-7} & j_{6}& \qhx^{5}\qlx\qlz^{2} & -5.92\times10^{-6}\\

f_{0}& \qhx\qlx & 0.0018 & j_{7}& \qhx^{4}\qlx^{2}\qlz^{2} & 5.2\times10^{-7}\\
f_{1}& \qhx^{3}\qlx & 0.05605 & j_{8}& \qhx^{3}\qlx^{3}\qlz^{2} & -2.0\times10^{-7}\\
f_{2}& \qhx^{2}\qlx^{2} & 0.02934 & j_{9}& \qhx^{2}\qlx^{4}\qlz^{2} & -6.8\times10^{-7}\\
f_{3}& \qhx\qlx^{3} & 0.0068 & j_{10}& \qhx\qlx^{5}\qlz^{2} & 1.0\times10^{-7}\\
f_{4}& \qhx^{5}\qlx & 6.03\times10^{-4} & j_{11}& \qhx^{3}\qlx\qlz^{4} & -2.18\times10^{-6}\\
f_{5}& \qhx^{4}\qlx^{2} & 1.54\times10^{-4} & j_{12}& \qhx^{2}\qlx^{2}\qlz^{4} & -2.94\times10^{-6}\\
f_{6}& \qhx^{3}\qlx^{3} & -6.83\times10^{-5} & j_{13}& \qhx\qlx^{3}\qlz^{4} & -6.3\times10^{-7}\\
f_{7}& \qhx^{2}\qlx^{4} & -8.83\times10^{-5} & j_{14}& \qhx\qlx\qlz^{6} & 7.4\times10^{-6}\\
f_{8}& \qhx\qlx^{5} & -1.62\times10^{-5}\\
\end{tabular}
\end{ruledtabular}
\end{table*}

\cleardoublepage

\bibliography{article}

\providecommand{\noopsort}[1]{}\providecommand{\singleletter}[1]{#1}%
\begin{thebibliography}{44}%
\makeatletter
\providecommand \@ifxundefined [1]{%
 \@ifx{#1\undefined}
}%
\providecommand \@ifnum [1]{%
 \ifnum #1\expandafter \@firstoftwo
 \else \expandafter \@secondoftwo
 \fi
}%
\providecommand \@ifx [1]{%
 \ifx #1\expandafter \@firstoftwo
 \else \expandafter \@secondoftwo
 \fi
}%
\providecommand \natexlab [1]{#1}%
\providecommand \enquote  [1]{``#1''}%
\providecommand \bibnamefont  [1]{#1}%
\providecommand \bibfnamefont [1]{#1}%
\providecommand \citenamefont [1]{#1}%
\providecommand \href@noop [0]{\@secondoftwo}%
\providecommand \href [0]{\begingroup \@sanitize@url \@href}%
\providecommand \@href[1]{\@@startlink{#1}\@@href}%
\providecommand \@@href[1]{\endgroup#1\@@endlink}%
\providecommand \@sanitize@url [0]{\catcode `\\12\catcode `\$12\catcode
  `\&12\catcode `\#12\catcode `\^12\catcode `\_12\catcode `\%12\relax}%
\providecommand \@@startlink[1]{}%
\providecommand \@@endlink[0]{}%
\providecommand \url  [0]{\begingroup\@sanitize@url \@url }%
\providecommand \@url [1]{\endgroup\@href {#1}{\urlprefix }}%
\providecommand \urlprefix  [0]{URL }%
\providecommand \Eprint [0]{\href }%
\providecommand \doibase [0]{https://doi.org/}%
\providecommand \selectlanguage [0]{\@gobble}%
\providecommand \bibinfo  [0]{\@secondoftwo}%
\providecommand \bibfield  [0]{\@secondoftwo}%
\providecommand \translation [1]{[#1]}%
\providecommand \BibitemOpen [0]{}%
\providecommand \bibitemStop [0]{}%
\providecommand \bibitemNoStop [0]{.\EOS\space}%
\providecommand \EOS [0]{\spacefactor3000\relax}%
\providecommand \BibitemShut  [1]{\csname bibitem#1\endcsname}%
\let\auto@bib@innerbib\@empty
\bibitem [{\citenamefont {Mankowsky}\ \emph {et~al.}(2016)\citenamefont
  {Mankowsky}, \citenamefont {F{\"o}rst},\ and\ \citenamefont
  {Cavalleri}}]{Mankowsky2016}%
  \BibitemOpen
  \bibfield  {author} {\bibinfo {author} {\bibfnamefont {R.}~\bibnamefont
  {Mankowsky}}, \bibinfo {author} {\bibfnamefont {M.}~\bibnamefont
  {F{\"o}rst}},\ and\ \bibinfo {author} {\bibfnamefont {A.}~\bibnamefont
  {Cavalleri}},\ }\href@noop {} {\bibfield  {journal} {\bibinfo  {journal}
  {Reports on Progress in Physics}\ }\textbf {\bibinfo {volume} {79}},\
  \bibinfo {pages} {064503} (\bibinfo {year} {2016})}\BibitemShut {NoStop}%
\bibitem [{\citenamefont {Sal{\'e}n}\ \emph {et~al.}(2019)\citenamefont
  {Sal{\'e}n}, \citenamefont {Basini}, \citenamefont {Bonetti}, \citenamefont
  {Hebling}, \citenamefont {Krasilnikov}, \citenamefont {Nikitin},
  \citenamefont {Shamuilov}, \citenamefont {Tibai}, \citenamefont
  {Zhaunerchyk},\ and\ \citenamefont {Goryashko}}]{Salen2019}%
  \BibitemOpen
  \bibfield  {author} {\bibinfo {author} {\bibfnamefont {P.}~\bibnamefont
  {Sal{\'e}n}}, \bibinfo {author} {\bibfnamefont {M.}~\bibnamefont {Basini}},
  \bibinfo {author} {\bibfnamefont {S.}~\bibnamefont {Bonetti}}, \bibinfo
  {author} {\bibfnamefont {J.}~\bibnamefont {Hebling}}, \bibinfo {author}
  {\bibfnamefont {M.}~\bibnamefont {Krasilnikov}}, \bibinfo {author}
  {\bibfnamefont {A.~Y.}\ \bibnamefont {Nikitin}}, \bibinfo {author}
  {\bibfnamefont {G.}~\bibnamefont {Shamuilov}}, \bibinfo {author}
  {\bibfnamefont {Z.}~\bibnamefont {Tibai}}, \bibinfo {author} {\bibfnamefont
  {V.}~\bibnamefont {Zhaunerchyk}},\ and\ \bibinfo {author} {\bibfnamefont
  {V.}~\bibnamefont {Goryashko}},\ }\href@noop {} {\bibfield  {journal}
  {\bibinfo  {journal} {Physics reports}\ }\textbf {\bibinfo {volume} {836}},\
  \bibinfo {pages} {1} (\bibinfo {year} {2019})}\BibitemShut {NoStop}%
\bibitem [{\citenamefont {Subedi}(2021)}]{Subedi2021}%
  \BibitemOpen
  \bibfield  {author} {\bibinfo {author} {\bibfnamefont {A.}~\bibnamefont
  {Subedi}},\ }\href {https://doi.org/10.5802/crphys.44} {\bibfield  {journal}
  {\bibinfo  {journal} {Comptes Rendus. Physique}\ }\textbf {\bibinfo {volume}
  {22}},\ \bibinfo {pages} {161} (\bibinfo {year} {2021})}\BibitemShut
  {NoStop}%
\bibitem [{\citenamefont {F{\"o}rst}\ \emph {et~al.}(2011)\citenamefont
  {F{\"o}rst}, \citenamefont {Manzoni}, \citenamefont {Kaiser}, \citenamefont
  {Tomioka}, \citenamefont {Tokura}, \citenamefont {Merlin},\ and\
  \citenamefont {Cavalleri}}]{Forst2011}%
  \BibitemOpen
  \bibfield  {author} {\bibinfo {author} {\bibfnamefont {M.}~\bibnamefont
  {F{\"o}rst}}, \bibinfo {author} {\bibfnamefont {C.}~\bibnamefont {Manzoni}},
  \bibinfo {author} {\bibfnamefont {S.}~\bibnamefont {Kaiser}}, \bibinfo
  {author} {\bibfnamefont {Y.}~\bibnamefont {Tomioka}}, \bibinfo {author}
  {\bibfnamefont {Y.}~\bibnamefont {Tokura}}, \bibinfo {author} {\bibfnamefont
  {R.}~\bibnamefont {Merlin}},\ and\ \bibinfo {author} {\bibfnamefont
  {A.}~\bibnamefont {Cavalleri}},\ }\href {https://doi.org/10.1038/nphys2055}
  {\bibfield  {journal} {\bibinfo  {journal} {Nature Physics}\ }\textbf
  {\bibinfo {volume} {7}},\ \bibinfo {pages} {854} (\bibinfo {year}
  {2011})}\BibitemShut {NoStop}%
\bibitem [{\citenamefont {Subedi}(2015)}]{Subedi2015}%
  \BibitemOpen
  \bibfield  {author} {\bibinfo {author} {\bibfnamefont {A.}~\bibnamefont
  {Subedi}},\ }\href {https://doi.org/10.1103/PhysRevB.92.214303} {\bibfield
  {journal} {\bibinfo  {journal} {Phys. Rev. B}\ }\textbf {\bibinfo {volume}
  {92}},\ \bibinfo {pages} {214303} (\bibinfo {year} {2015})}\BibitemShut
  {NoStop}%
\bibitem [{\citenamefont {Mankowsky}\ \emph {et~al.}(2017)\citenamefont
  {Mankowsky}, \citenamefont {von Hoegen}, \citenamefont {F\"orst},\ and\
  \citenamefont {Cavalleri}}]{Mankowsky2017}%
  \BibitemOpen
  \bibfield  {author} {\bibinfo {author} {\bibfnamefont {R.}~\bibnamefont
  {Mankowsky}}, \bibinfo {author} {\bibfnamefont {A.}~\bibnamefont {von
  Hoegen}}, \bibinfo {author} {\bibfnamefont {M.}~\bibnamefont {F\"orst}},\
  and\ \bibinfo {author} {\bibfnamefont {A.}~\bibnamefont {Cavalleri}},\ }\href
  {https://doi.org/10.1103/PhysRevLett.118.197601} {\bibfield  {journal}
  {\bibinfo  {journal} {Phys. Rev. Lett.}\ }\textbf {\bibinfo {volume} {118}},\
  \bibinfo {pages} {197601} (\bibinfo {year} {2017})}\BibitemShut {NoStop}%
\bibitem [{\citenamefont {Henstridge}\ \emph {et~al.}(2022)\citenamefont
  {Henstridge}, \citenamefont {F{\"o}rst}, \citenamefont {Rowe}, \citenamefont
  {Fechner},\ and\ \citenamefont {Cavalleri}}]{Henstridge2022}%
  \BibitemOpen
  \bibfield  {author} {\bibinfo {author} {\bibfnamefont {M.}~\bibnamefont
  {Henstridge}}, \bibinfo {author} {\bibfnamefont {M.}~\bibnamefont
  {F{\"o}rst}}, \bibinfo {author} {\bibfnamefont {E.}~\bibnamefont {Rowe}},
  \bibinfo {author} {\bibfnamefont {M.}~\bibnamefont {Fechner}},\ and\ \bibinfo
  {author} {\bibfnamefont {A.}~\bibnamefont {Cavalleri}},\ }\href@noop {}
  {\bibfield  {journal} {\bibinfo  {journal} {Nature Physics}\ }\textbf
  {\bibinfo {volume} {18}},\ \bibinfo {pages} {457} (\bibinfo {year}
  {2022})}\BibitemShut {NoStop}%
\bibitem [{\citenamefont {Wallis}\ and\ \citenamefont
  {Maradudin}(1971)}]{Wallis1971}%
  \BibitemOpen
  \bibfield  {author} {\bibinfo {author} {\bibfnamefont {R.}~\bibnamefont
  {Wallis}}\ and\ \bibinfo {author} {\bibfnamefont {A.}~\bibnamefont
  {Maradudin}},\ }\href@noop {} {\bibfield  {journal} {\bibinfo  {journal}
  {Physical Review B}\ }\textbf {\bibinfo {volume} {3}},\ \bibinfo {pages}
  {2063} (\bibinfo {year} {1971})}\BibitemShut {NoStop}%
\bibitem [{\citenamefont {Martin}\ and\ \citenamefont
  {Genzel}(1974)}]{Martin1974}%
  \BibitemOpen
  \bibfield  {author} {\bibinfo {author} {\bibfnamefont {T.}~\bibnamefont
  {Martin}}\ and\ \bibinfo {author} {\bibfnamefont {L.}~\bibnamefont
  {Genzel}},\ }\href@noop {} {\bibfield  {journal} {\bibinfo  {journal}
  {physica status solidi (b)}\ }\textbf {\bibinfo {volume} {61}},\ \bibinfo
  {pages} {493} (\bibinfo {year} {1974})}\BibitemShut {NoStop}%
\bibitem [{\citenamefont {Subedi}\ \emph {et~al.}(2014)\citenamefont {Subedi},
  \citenamefont {Cavalleri},\ and\ \citenamefont {Georges}}]{Subedi2014}%
  \BibitemOpen
  \bibfield  {author} {\bibinfo {author} {\bibfnamefont {A.}~\bibnamefont
  {Subedi}}, \bibinfo {author} {\bibfnamefont {A.}~\bibnamefont {Cavalleri}},\
  and\ \bibinfo {author} {\bibfnamefont {A.}~\bibnamefont {Georges}},\ }\href
  {https://doi.org/10.1103/PhysRevB.89.220301} {\bibfield  {journal} {\bibinfo
  {journal} {Phys. Rev. B}\ }\textbf {\bibinfo {volume} {89}},\ \bibinfo
  {pages} {220301} (\bibinfo {year} {2014})}\BibitemShut {NoStop}%
\bibitem [{\citenamefont {Subedi}(2017)}]{Subedi2017}%
  \BibitemOpen
  \bibfield  {author} {\bibinfo {author} {\bibfnamefont {A.}~\bibnamefont
  {Subedi}},\ }\href {https://doi.org/10.1103/PhysRevB.95.134113} {\bibfield
  {journal} {\bibinfo  {journal} {Phys. Rev. B}\ }\textbf {\bibinfo {volume}
  {95}},\ \bibinfo {pages} {134113} (\bibinfo {year} {2017})}\BibitemShut
  {NoStop}%
\bibitem [{\citenamefont {Radaelli}(2018)}]{Radaelli2018}%
  \BibitemOpen
  \bibfield  {author} {\bibinfo {author} {\bibfnamefont {P.~G.}\ \bibnamefont
  {Radaelli}},\ }\href@noop {} {\bibfield  {journal} {\bibinfo  {journal}
  {Physical Review B}\ }\textbf {\bibinfo {volume} {97}},\ \bibinfo {pages}
  {085145} (\bibinfo {year} {2018})}\BibitemShut {NoStop}%
\bibitem [{\citenamefont {Juraschek}\ \emph
  {et~al.}(2017{\natexlab{a}})\citenamefont {Juraschek}, \citenamefont
  {Fechner},\ and\ \citenamefont {Spaldin}}]{Juraschek2017}%
  \BibitemOpen
  \bibfield  {author} {\bibinfo {author} {\bibfnamefont {D.~M.}\ \bibnamefont
  {Juraschek}}, \bibinfo {author} {\bibfnamefont {M.}~\bibnamefont {Fechner}},\
  and\ \bibinfo {author} {\bibfnamefont {N.~A.}\ \bibnamefont {Spaldin}},\
  }\href@noop {} {\bibfield  {journal} {\bibinfo  {journal} {Physical review
  letters}\ }\textbf {\bibinfo {volume} {118}},\ \bibinfo {pages} {054101}
  (\bibinfo {year} {2017}{\natexlab{a}})}\BibitemShut {NoStop}%
\bibitem [{\citenamefont {Fechner}\ and\ \citenamefont
  {Spaldin}(2016)}]{Fechner2016}%
  \BibitemOpen
  \bibfield  {author} {\bibinfo {author} {\bibfnamefont {M.}~\bibnamefont
  {Fechner}}\ and\ \bibinfo {author} {\bibfnamefont {N.~A.}\ \bibnamefont
  {Spaldin}},\ }\href@noop {} {\bibfield  {journal} {\bibinfo  {journal}
  {Physical Review B}\ }\textbf {\bibinfo {volume} {94}},\ \bibinfo {pages}
  {134307} (\bibinfo {year} {2016})}\BibitemShut {NoStop}%
\bibitem [{\citenamefont {Gu}\ and\ \citenamefont {Rondinelli}(2016)}]{Gu2016}%
  \BibitemOpen
  \bibfield  {author} {\bibinfo {author} {\bibfnamefont {M.}~\bibnamefont
  {Gu}}\ and\ \bibinfo {author} {\bibfnamefont {J.~M.}\ \bibnamefont
  {Rondinelli}},\ }\href@noop {} {\bibfield  {journal} {\bibinfo  {journal}
  {Scientific reports}\ }\textbf {\bibinfo {volume} {6}},\ \bibinfo {pages} {1}
  (\bibinfo {year} {2016})}\BibitemShut {NoStop}%
\bibitem [{\citenamefont {Gu}\ and\ \citenamefont {Rondinelli}(2017)}]{Gu2017}%
  \BibitemOpen
  \bibfield  {author} {\bibinfo {author} {\bibfnamefont {M.}~\bibnamefont
  {Gu}}\ and\ \bibinfo {author} {\bibfnamefont {J.~M.}\ \bibnamefont
  {Rondinelli}},\ }\href@noop {} {\bibfield  {journal} {\bibinfo  {journal}
  {Physical Review B}\ }\textbf {\bibinfo {volume} {95}},\ \bibinfo {pages}
  {024109} (\bibinfo {year} {2017})}\BibitemShut {NoStop}%
\bibitem [{\citenamefont {Juraschek}\ \emph
  {et~al.}(2017{\natexlab{b}})\citenamefont {Juraschek}, \citenamefont
  {Fechner}, \citenamefont {Balatsky},\ and\ \citenamefont
  {Spaldin}}]{Juraschek2017b}%
  \BibitemOpen
  \bibfield  {author} {\bibinfo {author} {\bibfnamefont {D.~M.}\ \bibnamefont
  {Juraschek}}, \bibinfo {author} {\bibfnamefont {M.}~\bibnamefont {Fechner}},
  \bibinfo {author} {\bibfnamefont {A.~V.}\ \bibnamefont {Balatsky}},\ and\
  \bibinfo {author} {\bibfnamefont {N.~A.}\ \bibnamefont {Spaldin}},\
  }\href@noop {} {\bibfield  {journal} {\bibinfo  {journal} {Physical Review
  Materials}\ }\textbf {\bibinfo {volume} {1}},\ \bibinfo {pages} {014401}
  (\bibinfo {year} {2017}{\natexlab{b}})}\BibitemShut {NoStop}%
\bibitem [{\citenamefont {Fechner}\ \emph {et~al.}(2018)\citenamefont
  {Fechner}, \citenamefont {Sukhov}, \citenamefont {Chotorlishvili},
  \citenamefont {Kenel}, \citenamefont {Berakdar},\ and\ \citenamefont
  {Spaldin}}]{Fechner2018}%
  \BibitemOpen
  \bibfield  {author} {\bibinfo {author} {\bibfnamefont {M.}~\bibnamefont
  {Fechner}}, \bibinfo {author} {\bibfnamefont {A.}~\bibnamefont {Sukhov}},
  \bibinfo {author} {\bibfnamefont {L.}~\bibnamefont {Chotorlishvili}},
  \bibinfo {author} {\bibfnamefont {C.}~\bibnamefont {Kenel}}, \bibinfo
  {author} {\bibfnamefont {J.}~\bibnamefont {Berakdar}},\ and\ \bibinfo
  {author} {\bibfnamefont {N.}~\bibnamefont {Spaldin}},\ }\href@noop {}
  {\bibfield  {journal} {\bibinfo  {journal} {Physical review materials}\
  }\textbf {\bibinfo {volume} {2}},\ \bibinfo {pages} {064401} (\bibinfo {year}
  {2018})}\BibitemShut {NoStop}%
\bibitem [{\citenamefont {Gu}\ and\ \citenamefont {Rondinelli}(2018)}]{Gu2018}%
  \BibitemOpen
  \bibfield  {author} {\bibinfo {author} {\bibfnamefont {M.}~\bibnamefont
  {Gu}}\ and\ \bibinfo {author} {\bibfnamefont {J.~M.}\ \bibnamefont
  {Rondinelli}},\ }\href@noop {} {\bibfield  {journal} {\bibinfo  {journal}
  {Physical Review B}\ }\textbf {\bibinfo {volume} {98}},\ \bibinfo {pages}
  {024102} (\bibinfo {year} {2018})}\BibitemShut {NoStop}%
\bibitem [{\citenamefont {Khalsa}\ and\ \citenamefont
  {Benedek}(2018)}]{Khalsa2018}%
  \BibitemOpen
  \bibfield  {author} {\bibinfo {author} {\bibfnamefont {G.}~\bibnamefont
  {Khalsa}}\ and\ \bibinfo {author} {\bibfnamefont {N.~A.}\ \bibnamefont
  {Benedek}},\ }\href@noop {} {\bibfield  {journal} {\bibinfo  {journal} {npj
  Quantum Materials}\ }\textbf {\bibinfo {volume} {3}},\ \bibinfo {pages} {1}
  (\bibinfo {year} {2018})}\BibitemShut {NoStop}%
\bibitem [{\citenamefont {Park}\ \emph {et~al.}(2019)\citenamefont {Park},
  \citenamefont {Yeu}, \citenamefont {Han}, \citenamefont {Hwang},\ and\
  \citenamefont {Choi}}]{Park2019}%
  \BibitemOpen
  \bibfield  {author} {\bibinfo {author} {\bibfnamefont {J.}~\bibnamefont
  {Park}}, \bibinfo {author} {\bibfnamefont {I.~W.}\ \bibnamefont {Yeu}},
  \bibinfo {author} {\bibfnamefont {G.}~\bibnamefont {Han}}, \bibinfo {author}
  {\bibfnamefont {C.~S.}\ \bibnamefont {Hwang}},\ and\ \bibinfo {author}
  {\bibfnamefont {J.-H.}\ \bibnamefont {Choi}},\ }\href@noop {} {\bibfield
  {journal} {\bibinfo  {journal} {Scientific Reports}\ }\textbf {\bibinfo
  {volume} {9}},\ \bibinfo {pages} {1} (\bibinfo {year} {2019})}\BibitemShut
  {NoStop}%
\bibitem [{\citenamefont {Juraschek}\ \emph {et~al.}(2021)\citenamefont
  {Juraschek}, \citenamefont {Neuman}, \citenamefont {Flick},\ and\
  \citenamefont {Narang}}]{Juraschek2021}%
  \BibitemOpen
  \bibfield  {author} {\bibinfo {author} {\bibfnamefont {D.~M.}\ \bibnamefont
  {Juraschek}}, \bibinfo {author} {\bibfnamefont {T.}~\bibnamefont {Neuman}},
  \bibinfo {author} {\bibfnamefont {J.}~\bibnamefont {Flick}},\ and\ \bibinfo
  {author} {\bibfnamefont {P.}~\bibnamefont {Narang}},\ }\href@noop {}
  {\bibfield  {journal} {\bibinfo  {journal} {Physical Review Research}\
  }\textbf {\bibinfo {volume} {3}},\ \bibinfo {pages} {L032046} (\bibinfo
  {year} {2021})}\BibitemShut {NoStop}%
\bibitem [{\citenamefont {Kaaret}\ \emph {et~al.}(2021)\citenamefont {Kaaret},
  \citenamefont {Khalsa},\ and\ \citenamefont {Benedek}}]{Kaaret2021}%
  \BibitemOpen
  \bibfield  {author} {\bibinfo {author} {\bibfnamefont {J.~Z.}\ \bibnamefont
  {Kaaret}}, \bibinfo {author} {\bibfnamefont {G.}~\bibnamefont {Khalsa}},\
  and\ \bibinfo {author} {\bibfnamefont {N.~A.}\ \bibnamefont {Benedek}},\
  }\href@noop {} {\bibfield  {journal} {\bibinfo  {journal} {Journal of
  Physics: Condensed Matter}\ }\textbf {\bibinfo {volume} {34}},\ \bibinfo
  {pages} {035402} (\bibinfo {year} {2021})}\BibitemShut {NoStop}%
\bibitem [{\citenamefont {Feng}\ \emph {et~al.}(2022)\citenamefont {Feng},
  \citenamefont {Han}, \citenamefont {Lan}, \citenamefont {Xu}, \citenamefont
  {Bi}, \citenamefont {Lin},\ and\ \citenamefont {Nan}}]{Feng2022}%
  \BibitemOpen
  \bibfield  {author} {\bibinfo {author} {\bibfnamefont {N.}~\bibnamefont
  {Feng}}, \bibinfo {author} {\bibfnamefont {J.}~\bibnamefont {Han}}, \bibinfo
  {author} {\bibfnamefont {C.}~\bibnamefont {Lan}}, \bibinfo {author}
  {\bibfnamefont {B.}~\bibnamefont {Xu}}, \bibinfo {author} {\bibfnamefont
  {K.}~\bibnamefont {Bi}}, \bibinfo {author} {\bibfnamefont {Y.}~\bibnamefont
  {Lin}},\ and\ \bibinfo {author} {\bibfnamefont {C.}~\bibnamefont {Nan}},\
  }\href@noop {} {\bibfield  {journal} {\bibinfo  {journal} {Physical Review
  B}\ }\textbf {\bibinfo {volume} {105}},\ \bibinfo {pages} {024304} (\bibinfo
  {year} {2022})}\BibitemShut {NoStop}%
\bibitem [{\citenamefont {Nova}\ \emph {et~al.}(2017)\citenamefont {Nova},
  \citenamefont {Cartella}, \citenamefont {Cantaluppi}, \citenamefont
  {F{\"o}rst}, \citenamefont {Bossini}, \citenamefont {Mikhaylovskiy},
  \citenamefont {Kimel}, \citenamefont {Merlin},\ and\ \citenamefont
  {Cavalleri}}]{Nova2017}%
  \BibitemOpen
  \bibfield  {author} {\bibinfo {author} {\bibfnamefont {T.~F.}\ \bibnamefont
  {Nova}}, \bibinfo {author} {\bibfnamefont {A.}~\bibnamefont {Cartella}},
  \bibinfo {author} {\bibfnamefont {A.}~\bibnamefont {Cantaluppi}}, \bibinfo
  {author} {\bibfnamefont {M.}~\bibnamefont {F{\"o}rst}}, \bibinfo {author}
  {\bibfnamefont {D.}~\bibnamefont {Bossini}}, \bibinfo {author} {\bibfnamefont
  {R.~V.}\ \bibnamefont {Mikhaylovskiy}}, \bibinfo {author} {\bibfnamefont
  {A.~V.}\ \bibnamefont {Kimel}}, \bibinfo {author} {\bibfnamefont
  {R.}~\bibnamefont {Merlin}},\ and\ \bibinfo {author} {\bibfnamefont
  {A.}~\bibnamefont {Cavalleri}},\ }\href@noop {} {\bibfield  {journal}
  {\bibinfo  {journal} {Nature Physics}\ }\textbf {\bibinfo {volume} {13}},\
  \bibinfo {pages} {132} (\bibinfo {year} {2017})}\BibitemShut {NoStop}%
\bibitem [{\citenamefont {Hortensius}\ \emph {et~al.}(2020)\citenamefont
  {Hortensius}, \citenamefont {Afanasiev}, \citenamefont {Sasani},
  \citenamefont {Bousquet},\ and\ \citenamefont {Caviglia}}]{Hortensius2020}%
  \BibitemOpen
  \bibfield  {author} {\bibinfo {author} {\bibfnamefont {J.}~\bibnamefont
  {Hortensius}}, \bibinfo {author} {\bibfnamefont {D.}~\bibnamefont
  {Afanasiev}}, \bibinfo {author} {\bibfnamefont {A.}~\bibnamefont {Sasani}},
  \bibinfo {author} {\bibfnamefont {E.}~\bibnamefont {Bousquet}},\ and\
  \bibinfo {author} {\bibfnamefont {A.}~\bibnamefont {Caviglia}},\ }\href@noop
  {} {\bibfield  {journal} {\bibinfo  {journal} {npj Quantum Materials}\
  }\textbf {\bibinfo {volume} {5}},\ \bibinfo {pages} {1} (\bibinfo {year}
  {2020})}\BibitemShut {NoStop}%
\bibitem [{\citenamefont {Neugebauer}\ \emph {et~al.}(2021)\citenamefont
  {Neugebauer}, \citenamefont {Juraschek}, \citenamefont {Savoini},
  \citenamefont {Engeler}, \citenamefont {Boie}, \citenamefont {Abreu},
  \citenamefont {Spaldin},\ and\ \citenamefont {Johnson}}]{Neugebauer2021}%
  \BibitemOpen
  \bibfield  {author} {\bibinfo {author} {\bibfnamefont {M.~J.}\ \bibnamefont
  {Neugebauer}}, \bibinfo {author} {\bibfnamefont {D.~M.}\ \bibnamefont
  {Juraschek}}, \bibinfo {author} {\bibfnamefont {M.}~\bibnamefont {Savoini}},
  \bibinfo {author} {\bibfnamefont {P.}~\bibnamefont {Engeler}}, \bibinfo
  {author} {\bibfnamefont {L.}~\bibnamefont {Boie}}, \bibinfo {author}
  {\bibfnamefont {E.}~\bibnamefont {Abreu}}, \bibinfo {author} {\bibfnamefont
  {N.~A.}\ \bibnamefont {Spaldin}},\ and\ \bibinfo {author} {\bibfnamefont
  {S.~L.}\ \bibnamefont {Johnson}},\ }\href@noop {} {\bibfield  {journal}
  {\bibinfo  {journal} {Physical Review Research}\ }\textbf {\bibinfo {volume}
  {3}},\ \bibinfo {pages} {013126} (\bibinfo {year} {2021})}\BibitemShut
  {NoStop}%
\bibitem [{\citenamefont {Afanasiev}\ \emph {et~al.}(2021)\citenamefont
  {Afanasiev}, \citenamefont {Hortensius}, \citenamefont {Ivanov},
  \citenamefont {Sasani}, \citenamefont {Bousquet}, \citenamefont {Blanter},
  \citenamefont {Mikhaylovskiy}, \citenamefont {Kimel},\ and\ \citenamefont
  {Caviglia}}]{Afanasiev2021}%
  \BibitemOpen
  \bibfield  {author} {\bibinfo {author} {\bibfnamefont {D.}~\bibnamefont
  {Afanasiev}}, \bibinfo {author} {\bibfnamefont {J.}~\bibnamefont
  {Hortensius}}, \bibinfo {author} {\bibfnamefont {B.}~\bibnamefont {Ivanov}},
  \bibinfo {author} {\bibfnamefont {A.}~\bibnamefont {Sasani}}, \bibinfo
  {author} {\bibfnamefont {E.}~\bibnamefont {Bousquet}}, \bibinfo {author}
  {\bibfnamefont {Y.}~\bibnamefont {Blanter}}, \bibinfo {author} {\bibfnamefont
  {R.}~\bibnamefont {Mikhaylovskiy}}, \bibinfo {author} {\bibfnamefont
  {A.}~\bibnamefont {Kimel}},\ and\ \bibinfo {author} {\bibfnamefont
  {A.}~\bibnamefont {Caviglia}},\ }\href@noop {} {\bibfield  {journal}
  {\bibinfo  {journal} {Nature materials}\ }\textbf {\bibinfo {volume} {20}},\
  \bibinfo {pages} {607} (\bibinfo {year} {2021})}\BibitemShut {NoStop}%
\bibitem [{\citenamefont {Disa}\ \emph {et~al.}(2020)\citenamefont {Disa},
  \citenamefont {Fechner}, \citenamefont {Nova}, \citenamefont {Liu},
  \citenamefont {F{\"o}rst}, \citenamefont {Prabhakaran}, \citenamefont
  {Radaelli},\ and\ \citenamefont {Cavalleri}}]{Disa2020}%
  \BibitemOpen
  \bibfield  {author} {\bibinfo {author} {\bibfnamefont {A.~S.}\ \bibnamefont
  {Disa}}, \bibinfo {author} {\bibfnamefont {M.}~\bibnamefont {Fechner}},
  \bibinfo {author} {\bibfnamefont {T.~F.}\ \bibnamefont {Nova}}, \bibinfo
  {author} {\bibfnamefont {B.}~\bibnamefont {Liu}}, \bibinfo {author}
  {\bibfnamefont {M.}~\bibnamefont {F{\"o}rst}}, \bibinfo {author}
  {\bibfnamefont {D.}~\bibnamefont {Prabhakaran}}, \bibinfo {author}
  {\bibfnamefont {P.~G.}\ \bibnamefont {Radaelli}},\ and\ \bibinfo {author}
  {\bibfnamefont {A.}~\bibnamefont {Cavalleri}},\ }\href@noop {} {\bibfield
  {journal} {\bibinfo  {journal} {Nature Physics}\ }\textbf {\bibinfo {volume}
  {16}},\ \bibinfo {pages} {937} (\bibinfo {year} {2020})}\BibitemShut
  {NoStop}%
\bibitem [{\citenamefont {Melnikov}\ \emph {et~al.}(2020)\citenamefont
  {Melnikov}, \citenamefont {Selivanov},\ and\ \citenamefont
  {Chekalin}}]{Melnikov2020}%
  \BibitemOpen
  \bibfield  {author} {\bibinfo {author} {\bibfnamefont {A.}~\bibnamefont
  {Melnikov}}, \bibinfo {author} {\bibfnamefont {Y.~G.}\ \bibnamefont
  {Selivanov}},\ and\ \bibinfo {author} {\bibfnamefont {S.}~\bibnamefont
  {Chekalin}},\ }\href@noop {} {\bibfield  {journal} {\bibinfo  {journal}
  {Physical Review B}\ }\textbf {\bibinfo {volume} {102}},\ \bibinfo {pages}
  {224301} (\bibinfo {year} {2020})}\BibitemShut {NoStop}%
\bibitem [{\citenamefont {Stupakiewicz}\ \emph {et~al.}(2021)\citenamefont
  {Stupakiewicz}, \citenamefont {Davies}, \citenamefont {Szerenos},
  \citenamefont {Afanasiev}, \citenamefont {Rabinovich}, \citenamefont {Boris},
  \citenamefont {Caviglia}, \citenamefont {Kimel},\ and\ \citenamefont
  {Kirilyuk}}]{Stupakiewicz2021}%
  \BibitemOpen
  \bibfield  {author} {\bibinfo {author} {\bibfnamefont {A.}~\bibnamefont
  {Stupakiewicz}}, \bibinfo {author} {\bibfnamefont {C.}~\bibnamefont
  {Davies}}, \bibinfo {author} {\bibfnamefont {K.}~\bibnamefont {Szerenos}},
  \bibinfo {author} {\bibfnamefont {D.}~\bibnamefont {Afanasiev}}, \bibinfo
  {author} {\bibfnamefont {K.}~\bibnamefont {Rabinovich}}, \bibinfo {author}
  {\bibfnamefont {A.}~\bibnamefont {Boris}}, \bibinfo {author} {\bibfnamefont
  {A.}~\bibnamefont {Caviglia}}, \bibinfo {author} {\bibfnamefont
  {A.}~\bibnamefont {Kimel}},\ and\ \bibinfo {author} {\bibfnamefont
  {A.}~\bibnamefont {Kirilyuk}},\ }\href@noop {} {\bibfield  {journal}
  {\bibinfo  {journal} {Nature Physics}\ }\textbf {\bibinfo {volume} {17}},\
  \bibinfo {pages} {489} (\bibinfo {year} {2021})}\BibitemShut {NoStop}%
\bibitem [{\citenamefont {Disa}\ \emph {et~al.}(2021)\citenamefont {Disa},
  \citenamefont {Curtis}, \citenamefont {Fechner}, \citenamefont {Liu},
  \citenamefont {von Hoegen}, \citenamefont {F{\"o}rst}, \citenamefont {Nova},
  \citenamefont {Narang}, \citenamefont {Maljuk}, \citenamefont {Boris} \emph
  {et~al.}}]{Disa2021}%
  \BibitemOpen
  \bibfield  {author} {\bibinfo {author} {\bibfnamefont {A.}~\bibnamefont
  {Disa}}, \bibinfo {author} {\bibfnamefont {J.}~\bibnamefont {Curtis}},
  \bibinfo {author} {\bibfnamefont {M.}~\bibnamefont {Fechner}}, \bibinfo
  {author} {\bibfnamefont {A.}~\bibnamefont {Liu}}, \bibinfo {author}
  {\bibfnamefont {A.}~\bibnamefont {von Hoegen}}, \bibinfo {author}
  {\bibfnamefont {M.}~\bibnamefont {F{\"o}rst}}, \bibinfo {author}
  {\bibfnamefont {T.}~\bibnamefont {Nova}}, \bibinfo {author} {\bibfnamefont
  {P.}~\bibnamefont {Narang}}, \bibinfo {author} {\bibfnamefont
  {A.}~\bibnamefont {Maljuk}}, \bibinfo {author} {\bibfnamefont
  {A.}~\bibnamefont {Boris}}, \emph {et~al.},\ }\href@noop {} {\bibfield
  {journal} {\bibinfo  {journal} {arXiv preprint arXiv:2111.13622}\ } (\bibinfo
  {year} {2021})}\BibitemShut {NoStop}%
\bibitem [{\citenamefont {Nilsen}\ and\ \citenamefont
  {Skinner}(1967)}]{Nilsen1967}%
  \BibitemOpen
  \bibfield  {author} {\bibinfo {author} {\bibfnamefont {W.~G.}\ \bibnamefont
  {Nilsen}}\ and\ \bibinfo {author} {\bibfnamefont {J.~G.}\ \bibnamefont
  {Skinner}},\ }\href {https://doi.org/10.1063/1.1712096} {\bibfield  {journal}
  {\bibinfo  {journal} {The Journal of Chemical Physics}\ }\textbf {\bibinfo
  {volume} {47}},\ \bibinfo {pages} {1413} (\bibinfo {year} {1967})},\ \Eprint
  {https://arxiv.org/abs/https://doi.org/10.1063/1.1712096}
  {https://doi.org/10.1063/1.1712096} \BibitemShut {NoStop}%
\bibitem [{\citenamefont {Cartella}\ \emph {et~al.}(2018)\citenamefont
  {Cartella}, \citenamefont {Nova}, \citenamefont {Fechner}, \citenamefont
  {Merlin},\ and\ \citenamefont {Cavalleri}}]{Cartella2018}%
  \BibitemOpen
  \bibfield  {author} {\bibinfo {author} {\bibfnamefont {A.}~\bibnamefont
  {Cartella}}, \bibinfo {author} {\bibfnamefont {T.~F.}\ \bibnamefont {Nova}},
  \bibinfo {author} {\bibfnamefont {M.}~\bibnamefont {Fechner}}, \bibinfo
  {author} {\bibfnamefont {R.}~\bibnamefont {Merlin}},\ and\ \bibinfo {author}
  {\bibfnamefont {A.}~\bibnamefont {Cavalleri}},\ }\href
  {https://doi.org/10.1073/pnas.1809725115} {\bibfield  {journal} {\bibinfo
  {journal} {Proceedings of the National Academy of Sciences}\ }\textbf
  {\bibinfo {volume} {115}},\ \bibinfo {pages} {12148} (\bibinfo {year}
  {2018})},\ \Eprint
  {https://arxiv.org/abs/https://www.pnas.org/content/115/48/12148.full.pdf}
  {https://www.pnas.org/content/115/48/12148.full.pdf} \BibitemShut {NoStop}%
\bibitem [{\citenamefont {Giannozzi}\ \emph {et~al.}(2020)\citenamefont
  {Giannozzi}, \citenamefont {Baseggio}, \citenamefont {Bonfà}, \citenamefont
  {Brunato}, \citenamefont {Car}, \citenamefont {Carnimeo}, \citenamefont
  {Cavazzoni}, \citenamefont {de~Gironcoli}, \citenamefont {Delugas},
  \citenamefont {Ferrari~Ruffino}, \citenamefont {Ferretti}, \citenamefont
  {Marzari}, \citenamefont {Timrov}, \citenamefont {Urru},\ and\ \citenamefont
  {Baroni}}]{QE}%
  \BibitemOpen
  \bibfield  {author} {\bibinfo {author} {\bibfnamefont {P.}~\bibnamefont
  {Giannozzi}}, \bibinfo {author} {\bibfnamefont {O.}~\bibnamefont {Baseggio}},
  \bibinfo {author} {\bibfnamefont {P.}~\bibnamefont {Bonfà}}, \bibinfo
  {author} {\bibfnamefont {D.}~\bibnamefont {Brunato}}, \bibinfo {author}
  {\bibfnamefont {R.}~\bibnamefont {Car}}, \bibinfo {author} {\bibfnamefont
  {I.}~\bibnamefont {Carnimeo}}, \bibinfo {author} {\bibfnamefont
  {C.}~\bibnamefont {Cavazzoni}}, \bibinfo {author} {\bibfnamefont
  {S.}~\bibnamefont {de~Gironcoli}}, \bibinfo {author} {\bibfnamefont
  {P.}~\bibnamefont {Delugas}}, \bibinfo {author} {\bibfnamefont
  {F.}~\bibnamefont {Ferrari~Ruffino}}, \bibinfo {author} {\bibfnamefont
  {A.}~\bibnamefont {Ferretti}}, \bibinfo {author} {\bibfnamefont
  {N.}~\bibnamefont {Marzari}}, \bibinfo {author} {\bibfnamefont
  {I.}~\bibnamefont {Timrov}}, \bibinfo {author} {\bibfnamefont
  {A.}~\bibnamefont {Urru}},\ and\ \bibinfo {author} {\bibfnamefont
  {S.}~\bibnamefont {Baroni}},\ }\href {https://doi.org/10.1063/5.0005082}
  {\bibfield  {journal} {\bibinfo  {journal} {The Journal of Chemical Physics}\
  }\textbf {\bibinfo {volume} {152}},\ \bibinfo {pages} {154105} (\bibinfo
  {year} {2020})},\ \Eprint
  {https://arxiv.org/abs/https://doi.org/10.1063/5.0005082}
  {https://doi.org/10.1063/5.0005082} \BibitemShut {NoStop}%
\bibitem [{\citenamefont {Garrity}\ \emph {et~al.}(2014)\citenamefont
  {Garrity}, \citenamefont {Bennett}, \citenamefont {Rabe},\ and\ \citenamefont
  {Vanderbilt}}]{GBRV}%
  \BibitemOpen
  \bibfield  {author} {\bibinfo {author} {\bibfnamefont {K.~F.}\ \bibnamefont
  {Garrity}}, \bibinfo {author} {\bibfnamefont {J.~W.}\ \bibnamefont
  {Bennett}}, \bibinfo {author} {\bibfnamefont {K.~M.}\ \bibnamefont {Rabe}},\
  and\ \bibinfo {author} {\bibfnamefont {D.}~\bibnamefont {Vanderbilt}},\
  }\href {https://doi.org/https://doi.org/10.1016/j.commatsci.2013.08.053}
  {\bibfield  {journal} {\bibinfo  {journal} {Computational Materials Science}\
  }\textbf {\bibinfo {volume} {81}},\ \bibinfo {pages} {446} (\bibinfo {year}
  {2014})}\BibitemShut {NoStop}%
\bibitem [{\citenamefont {Perdew}\ \emph {et~al.}(2008)\citenamefont {Perdew},
  \citenamefont {Ruzsinszky}, \citenamefont {Csonka}, \citenamefont {Vydrov},
  \citenamefont {Scuseria}, \citenamefont {Constantin}, \citenamefont {Zhou},\
  and\ \citenamefont {Burke}}]{PBEsol}%
  \BibitemOpen
  \bibfield  {author} {\bibinfo {author} {\bibfnamefont {J.~P.}\ \bibnamefont
  {Perdew}}, \bibinfo {author} {\bibfnamefont {A.}~\bibnamefont {Ruzsinszky}},
  \bibinfo {author} {\bibfnamefont {G.~I.}\ \bibnamefont {Csonka}}, \bibinfo
  {author} {\bibfnamefont {O.~A.}\ \bibnamefont {Vydrov}}, \bibinfo {author}
  {\bibfnamefont {G.~E.}\ \bibnamefont {Scuseria}}, \bibinfo {author}
  {\bibfnamefont {L.~A.}\ \bibnamefont {Constantin}}, \bibinfo {author}
  {\bibfnamefont {X.}~\bibnamefont {Zhou}},\ and\ \bibinfo {author}
  {\bibfnamefont {K.}~\bibnamefont {Burke}},\ }\href
  {https://doi.org/10.1103/PhysRevLett.100.136406} {\bibfield  {journal}
  {\bibinfo  {journal} {Phys. Rev. Lett.}\ }\textbf {\bibinfo {volume} {100}},\
  \bibinfo {pages} {136406} (\bibinfo {year} {2008})}\BibitemShut {NoStop}%
\bibitem [{\citenamefont {Verma}\ and\ \citenamefont
  {Jindal}(2009)}]{Verma2009}%
  \BibitemOpen
  \bibfield  {author} {\bibinfo {author} {\bibfnamefont {A.}~\bibnamefont
  {Verma}}\ and\ \bibinfo {author} {\bibfnamefont {V.}~\bibnamefont {Jindal}},\
  }\href {https://doi.org/https://doi.org/10.1016/j.jallcom.2009.06.001}
  {\bibfield  {journal} {\bibinfo  {journal} {Journal of Alloys and Compounds}\
  }\textbf {\bibinfo {volume} {485}},\ \bibinfo {pages} {514} (\bibinfo {year}
  {2009})}\BibitemShut {NoStop}%
\bibitem [{\citenamefont {Savrasov}\ \emph {et~al.}(1994)\citenamefont
  {Savrasov}, \citenamefont {Savrasov},\ and\ \citenamefont
  {Andersen}}]{Savrasov1994}%
  \BibitemOpen
  \bibfield  {author} {\bibinfo {author} {\bibfnamefont {S.~Y.}\ \bibnamefont
  {Savrasov}}, \bibinfo {author} {\bibfnamefont {D.~Y.}\ \bibnamefont
  {Savrasov}},\ and\ \bibinfo {author} {\bibfnamefont {O.~K.}\ \bibnamefont
  {Andersen}},\ }\href {https://doi.org/10.1103/PhysRevLett.72.372} {\bibfield
  {journal} {\bibinfo  {journal} {Phys. Rev. Lett.}\ }\textbf {\bibinfo
  {volume} {72}},\ \bibinfo {pages} {372} (\bibinfo {year} {1994})}\BibitemShut
  {NoStop}%
\bibitem [{\citenamefont {Bates}\ \emph {et~al.}(2022)\citenamefont {Bates},
  \citenamefont {Kornblith}, \citenamefont {Noack}, \citenamefont
  {Bouchet-Valat}, \citenamefont {Borregaard}, \citenamefont {Arslan},
  \citenamefont {White}, \citenamefont {Kleinschmidt}, \citenamefont {Lynch},
  \citenamefont {Dunning}, \citenamefont {Mogensen}, \citenamefont {Lendle},
  \citenamefont {Aluthge}, \citenamefont {pdeffebach}, \citenamefont {José
  Bayoán Santiago~Calderón}, \citenamefont {Born}, \citenamefont {Setzler},
  \citenamefont {DuBois}, \citenamefont {Quinn}, \citenamefont {Slámečka},
  \citenamefont {Bastide}, \citenamefont {Alday}, \citenamefont
  {Anthony~Blaom}, \citenamefont {König}, \citenamefont {Kamiński},
  \citenamefont {Caine}, \citenamefont {Lin},\ and\ \citenamefont
  {Karrasch}}]{GLM}%
  \BibitemOpen
  \bibfield  {author} {\bibinfo {author} {\bibfnamefont {D.}~\bibnamefont
  {Bates}}, \bibinfo {author} {\bibfnamefont {S.}~\bibnamefont {Kornblith}},
  \bibinfo {author} {\bibfnamefont {A.}~\bibnamefont {Noack}}, \bibinfo
  {author} {\bibfnamefont {M.}~\bibnamefont {Bouchet-Valat}}, \bibinfo {author}
  {\bibfnamefont {M.~K.}\ \bibnamefont {Borregaard}}, \bibinfo {author}
  {\bibfnamefont {A.}~\bibnamefont {Arslan}}, \bibinfo {author} {\bibfnamefont
  {J.~M.}\ \bibnamefont {White}}, \bibinfo {author} {\bibfnamefont
  {D.}~\bibnamefont {Kleinschmidt}}, \bibinfo {author} {\bibfnamefont
  {G.}~\bibnamefont {Lynch}}, \bibinfo {author} {\bibfnamefont
  {I.}~\bibnamefont {Dunning}}, \bibinfo {author} {\bibfnamefont {P.~K.}\
  \bibnamefont {Mogensen}}, \bibinfo {author} {\bibfnamefont {S.}~\bibnamefont
  {Lendle}}, \bibinfo {author} {\bibfnamefont {D.}~\bibnamefont {Aluthge}},
  \bibinfo {author} {\bibnamefont {pdeffebach}}, \bibinfo {author}
  {\bibfnamefont {P.}~\bibnamefont {José Bayoán Santiago~Calderón}},
  \bibinfo {author} {\bibfnamefont {B.}~\bibnamefont {Born}}, \bibinfo {author}
  {\bibfnamefont {B.}~\bibnamefont {Setzler}}, \bibinfo {author} {\bibfnamefont
  {C.}~\bibnamefont {DuBois}}, \bibinfo {author} {\bibfnamefont
  {J.}~\bibnamefont {Quinn}}, \bibinfo {author} {\bibfnamefont
  {O.}~\bibnamefont {Slámečka}}, \bibinfo {author} {\bibfnamefont
  {P.}~\bibnamefont {Bastide}}, \bibinfo {author} {\bibfnamefont
  {P.}~\bibnamefont {Alday}}, \bibinfo {author} {\bibfnamefont
  {P.}~\bibnamefont {Anthony~Blaom}}, \bibinfo {author} {\bibfnamefont
  {B.}~\bibnamefont {König}}, \bibinfo {author} {\bibfnamefont
  {B.}~\bibnamefont {Kamiński}}, \bibinfo {author} {\bibfnamefont
  {C.}~\bibnamefont {Caine}}, \bibinfo {author} {\bibfnamefont
  {D.}~\bibnamefont {Lin}},\ and\ \bibinfo {author} {\bibfnamefont
  {D.}~\bibnamefont {Karrasch}},\ }\href
  {https://doi.org/10.5281/zenodo.5823359} {\bibinfo {title}
  {Juliastats/glm.jl: v1.6.0}} (\bibinfo {year} {2022})\BibitemShut {NoStop}%
\bibitem [{\citenamefont {Souza}\ \emph {et~al.}(2002)\citenamefont {Souza},
  \citenamefont {\'I\~niguez},\ and\ \citenamefont {Vanderbilt}}]{Souza2002}%
  \BibitemOpen
  \bibfield  {author} {\bibinfo {author} {\bibfnamefont {I.}~\bibnamefont
  {Souza}}, \bibinfo {author} {\bibfnamefont {J.}~\bibnamefont {\'I\~niguez}},\
  and\ \bibinfo {author} {\bibfnamefont {D.}~\bibnamefont {Vanderbilt}},\
  }\href {https://doi.org/10.1103/PhysRevLett.89.117602} {\bibfield  {journal}
  {\bibinfo  {journal} {Phys. Rev. Lett.}\ }\textbf {\bibinfo {volume} {89}},\
  \bibinfo {pages} {117602} (\bibinfo {year} {2002})}\BibitemShut {NoStop}%
\bibitem [{\citenamefont {Bartels}\ \emph {et~al.}(2000)\citenamefont
  {Bartels}, \citenamefont {Dekorsy},\ and\ \citenamefont
  {Kurz}}]{Bartels2000}%
  \BibitemOpen
  \bibfield  {author} {\bibinfo {author} {\bibfnamefont {A.}~\bibnamefont
  {Bartels}}, \bibinfo {author} {\bibfnamefont {T.}~\bibnamefont {Dekorsy}},\
  and\ \bibinfo {author} {\bibfnamefont {H.}~\bibnamefont {Kurz}},\ }\href
  {https://doi.org/10.1103/PhysRevLett.84.2981} {\bibfield  {journal} {\bibinfo
   {journal} {Phys. Rev. Lett.}\ }\textbf {\bibinfo {volume} {84}},\ \bibinfo
  {pages} {2981} (\bibinfo {year} {2000})}\BibitemShut {NoStop}%
\bibitem [{\citenamefont {Rackauckas}\ and\ \citenamefont
  {Nie}(2017)}]{DifferentialEquations}%
  \BibitemOpen
  \bibfield  {author} {\bibinfo {author} {\bibfnamefont {C.}~\bibnamefont
  {Rackauckas}}\ and\ \bibinfo {author} {\bibfnamefont {Q.}~\bibnamefont
  {Nie}},\ }\href {https://doi.org/10.5334/jors.151} {\bibfield  {journal}
  {\bibinfo  {journal} {The Journal of Open Research Software}\ }\textbf
  {\bibinfo {volume} {5}} (\bibinfo {year} {2017})},\ \bibinfo {note} {exported
  from https://app.dimensions.ai on 2019/05/05}\BibitemShut {NoStop}%
\bibitem [{\citenamefont {Frigo}\ and\ \citenamefont {Johnson}(2005)}]{FFTW}%
  \BibitemOpen
  \bibfield  {author} {\bibinfo {author} {\bibfnamefont {M.}~\bibnamefont
  {Frigo}}\ and\ \bibinfo {author} {\bibfnamefont {S.~G.}\ \bibnamefont
  {Johnson}},\ }\href {https://doi.org/10.1109/JPROC.2004.840301} {\bibfield
  {journal} {\bibinfo  {journal} {Proceedings of the IEEE}\ }\textbf {\bibinfo
  {volume} {93}},\ \bibinfo {pages} {216} (\bibinfo {year} {2005})},\ \bibinfo
  {note} {special issue on ``Program Generation, Optimization, and Platform
  Adaptation''}\BibitemShut {NoStop}%
\end{thebibliography}%

\end{document}